\documentclass[preprint,12pt]{elsarticle}
\usepackage{natbib}
\usepackage{afterpage}
\usepackage{graphics}
\usepackage{amssymb}
\usepackage{longtable}
\newcommand{\bc}{\begin{center}}
\newcommand{\ec}{\end{center}}
\newcommand{\be}{\begin{equation}}
\newcommand{\ee}{\end{equation}}
\newcommand{\ba}{\begin{array}}
\newcommand{\ea}{\end{array}}
\newcommand{\bea}{\begin{eqnarray}}
\newcommand{\eea}{\end{eqnarray}}
\def\ga{\mathrel{\mathchoice {\vcenter{\offinterlineskip\halign{\hfil
$\displaystyle##$\hfil\cr>\cr\sim\cr}}}
{\vcenter{\offinterlineskip\halign{\hfil$\textstyle##$\hfil\cr
>\cr\sim\cr}}}
{\vcenter{\offinterlineskip\halign{\hfil$\scriptstyle##$\hfil\cr
>\cr\sim\cr}}}
{\vcenter{\offinterlineskip\halign{\hfil$\scriptscriptstyle##$\hfil\cr
>\cr\sim\cr}}}}}
\def\la{\mathrel{\mathchoice {\vcenter{\offinterlineskip\halign{\hfil
$\displaystyle##$\hfil\cr<\cr\sim\cr}}}
{\vcenter{\offinterlineskip\halign{\hfil$\textstyle##$\hfil\cr
<\cr\sim\cr}}}
{\vcenter{\offinterlineskip\halign{\hfil$\scriptstyle##$\hfil\cr
<\cr\sim\cr}}}
{\vcenter{\offinterlineskip\halign{\hfil$\scriptscriptstyle##$\hfil\cr
<\cr\sim\cr}}}}}
\def\mkm{{\mu}\rm{m}}

\def\is{interstellar }
\def\degr{\hbox{$^\circ$}}
\def\fdegr{\hbox{$.\!\!^\circ$}}
\def\fm{\hbox{$.\!\!^{\rm m}$}}

\journal{JQSRT}

\begin{document}

\begin{frontmatter}



\title{Interstellar extinction and interstellar polarization: old and new models}


\author{N.V.~Voshchinnikov}

\address{Sobolev Astronomical Institute, St.~Petersburg University,
Universitetskii prosp., 28, St.~Petersburg, 198504 Russia}

\ead{nvv@astro.spbu.ru}

\begin{abstract}
The review contains an analysis of the observed and model curves of
the interstellar extinction and polarization.
The observations mainly give information on dust in diffuse and
translucent interstellar clouds.
The features of various dust grain models including
spherical/non-spherical, homogeneous/inhomogeneous particles are discussed.
A special attention is devoted to the analysis of the grain size distributions,
alignment mechanisms
and magnetic field structure in \is clouds.
It is concluded  that the interpretation of interstellar
extinction and polarization is not yet complete.
\end{abstract}


\end{frontmatter}

\section{Introduction}

The properties of cosmic dust grains in various objects from comets
to distant galaxies are derived from
observations of interstellar extinction, \is polarization,
scattered radiation, infrared (IR) continuum emission  and IR features.
Modelling of these observations is aimed at estimates of the grain
size, chemical composition, shape, structure, and alignment.
The observed wavelength dependencies of \is extinction and polarization
({\rm interstellar extinction $A(\lambda)$
and polarization $P(\lambda)$ curves})
still remain the main  sources of information on dust in diffuse and
translucent \is clouds.

In this review, we discuss  observations
and modelling of the \is extinction and polarization curves.
Special attention is paid to various dust grain models.
For an extended consideration of the properties of dust
in different astronomical objects see \cite{w03,v04,witt04,henn09,henn10}.
Excellent historical reviews on dust astrophysics are given by
Dorschner \cite{dor10} and Li \cite{li05}.

\section{Extinction}

\subsection{Extinction curve: production and fitting}\label{obse}

The normalized extinction curves $A^{(n)}(\lambda^{-1})$ commonly
studied can be calculated as the ratio of
the colour excess of a star behind the dusty cloud $E(\lambda-V)$
to the colour excess $E(B-V)$
\begin{equation}
A^{(n)}(\lambda^{-1})\equiv \frac{E(\lambda-V)}{E(B-V)}=
\frac{A(\lambda)-A_V}{A_B-A_V}.
\label{norm}
\end{equation}
Here, $A_B$ and $A_V$ is the extinction in the B and V bands,
respectively. Sometimes,  the normalization of the extinction curve
on $A_V$ or extinction in another band is used.

In such a manner we can determine only the
``selective'' extinction (reddening), i.e. the difference of extinction
at two wavelengths. The absolute value of extinction can be found as
\begin{equation}
A_V=R_V E(B-V), \label{a_v}
\end{equation} 
where the coefficient $R_V$  is often evaluated from
observations in the visible and IR taking into account that
$A_{\lambda} \to 0$ when $\lambda \to \infty$.
From Eq.~(\ref{norm}), it follows that
\begin{equation}
R_V=\frac{A_V}{E(B-V)}=-\frac{E(\infty-V)}{E(B-V)}. \label{r_v}
\end{equation}
In the diffuse medium on average $R_V = 2.4-3.6$ \cite{fm07}.

\begin{figure}[htb]
\centerline{
\resizebox{7cm}{!}{\includegraphics{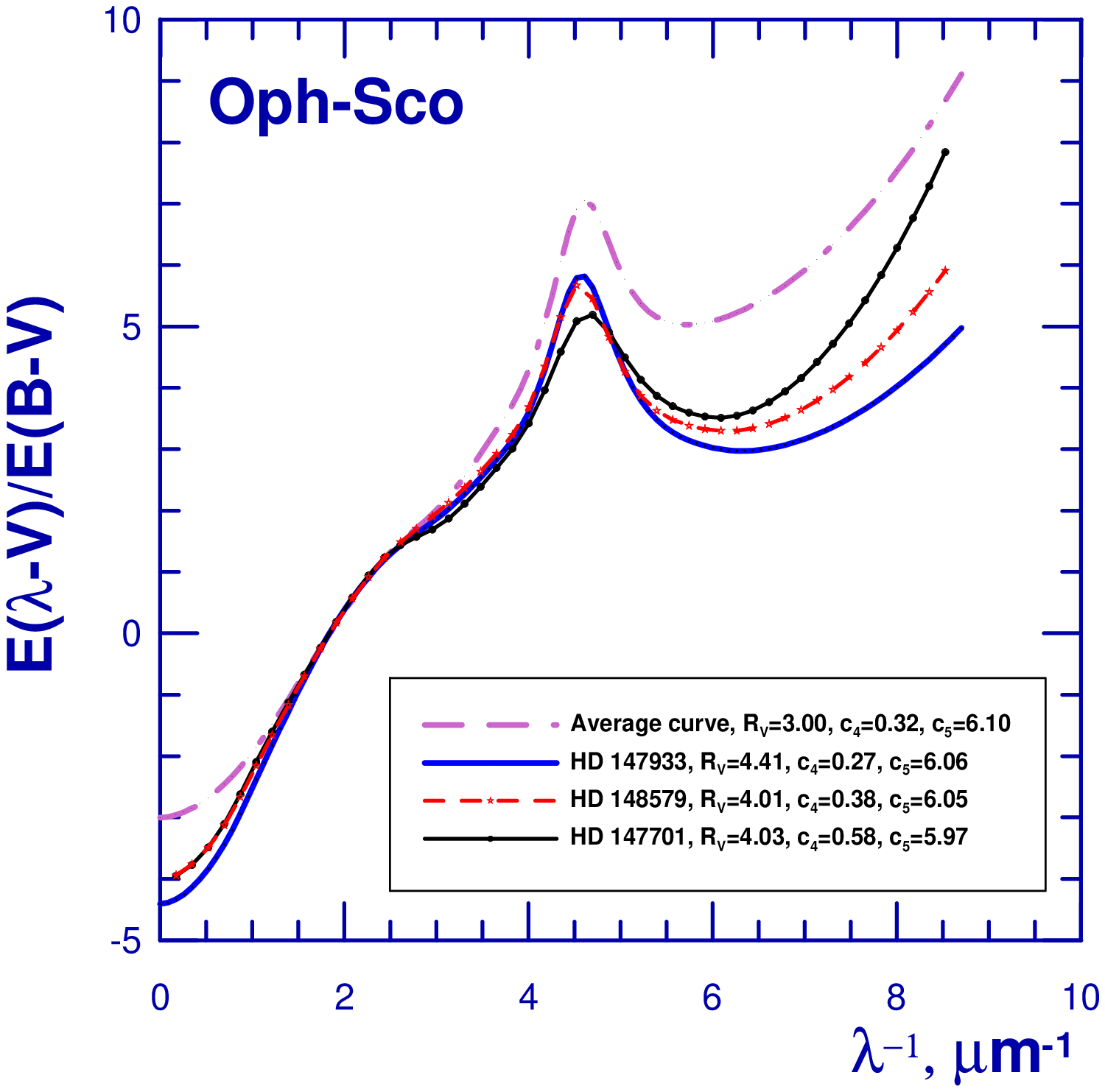}}
\resizebox{7cm}{!}{\includegraphics{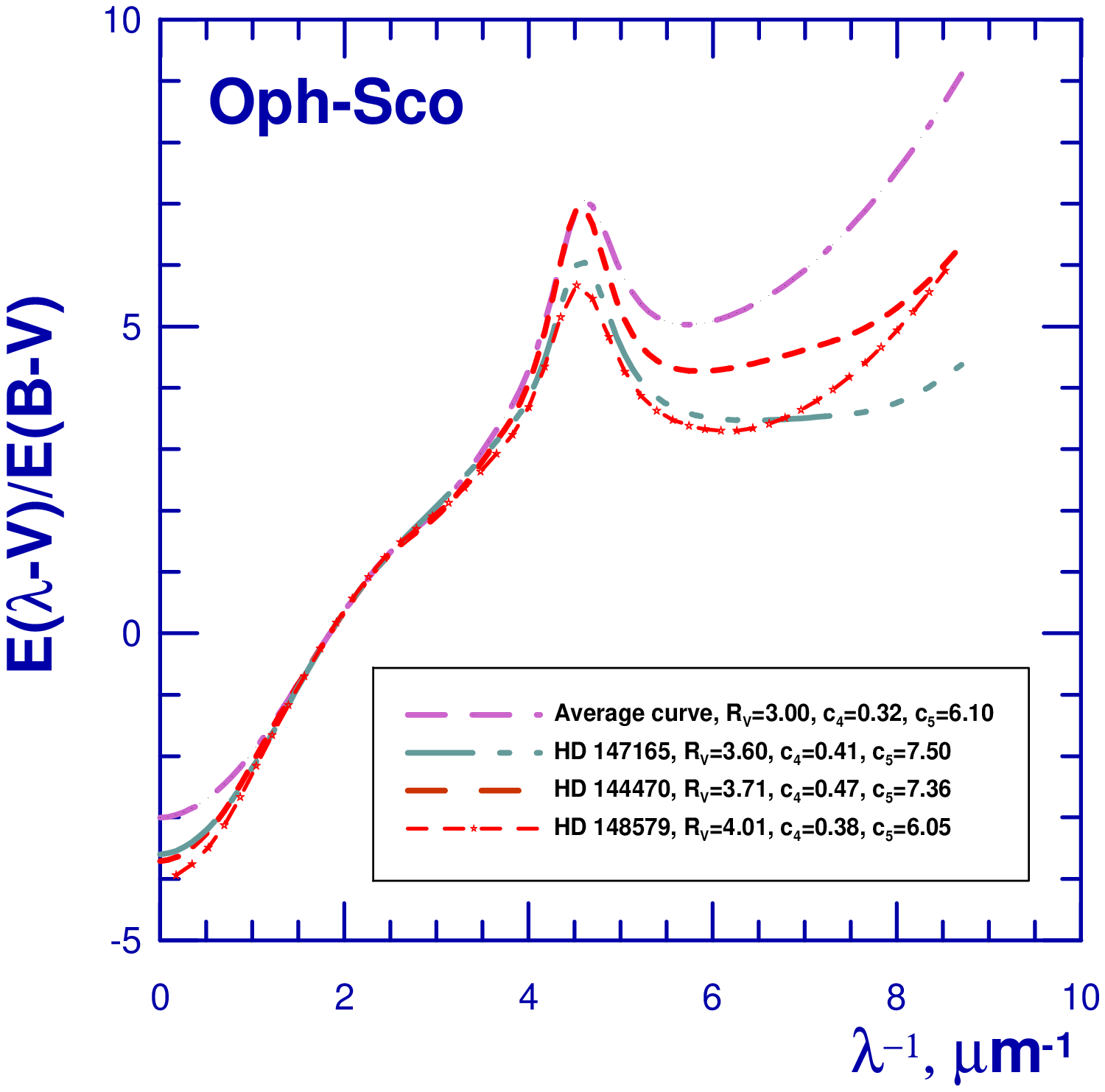}}
}
\caption
{The average extinction curve for 243 Galactic stars with $2.4<R_V<3.6$
and the extinction curves in
the direction of six stars in Sco-Oph \cite{fm07}.
The values of the coefficient $R_V$ and the UV fitting coefficients
$c_4$ and $c_5$ (Eq.~(\ref{fm_par})) are indicated in the legend.
The effect of variations of coefficients $c_4$ (left panel)
and $c_5$ (right panel) is illustrated. Adapted from \cite{fm07}.
}
\label{fext}
\end{figure}

In general, the  \is extinction curve has
a power law-like rise from the IR to the visible,
a prominent feature (bump) near $\lambda$~2175\,\AA, and  a
steep rise in the far-UV.
This rise is a manifestation of the very strong feature with
a maximum near $\lambda \approx 700$\,\AA\,  \cite{vi93}.
Figure~\ref{fext} shows the extinction
curve averaged over 243 Galactic B and late--O stars \cite{fm07}.
In the near IR-visible part of the spectrum, the distinction between the
extinction curves of different stars is rather small.
The IR extinction at wavelengths $\lambda=0.7$--$5\,\mkm$
was approximated by the power-law dependence:
$A(\lambda) \propto \lambda^{-\beta}$ with $\beta=-1.84$ \cite{marwh90}.
Fitzpatrick and Massa~\cite{fm09} found that the index $\beta$ was
$R_V$-dependent: $\beta > 2$ if $R_V<3$ and $\beta <1.5$ if $R_V>3$.
The mid-IR extinction at wavelengths $\lambda \ga 3\,\mkm$ measured
in the Galactic plane \cite{inde05} and toward the Galactic centre
\cite{fritz11} becomes grayer than the near-IR extinction.

In the UV region, the extinction curves differ  strongly \cite{fm07,val04,gor10}
(see also Fig.~\ref{fext}) demonstrating
that the mean curve in the UV obviously has little meaning.
The position of the UV bump center varies a little from star to star and
occurs at $\lambda_{0} = 2174 \pm  17$\,\AA~ or
$\lambda_0^{-1} = 4.599 \pm  0.012 \,\mkm^{-1}$.
The total half-width of the bump is $W = 0.992 \pm  0.058\, \mkm^{-1}$
that corresponds to $470 \pm  27$\,\AA \, \cite{fm86}.

Dorschner~\cite{dor73} was the first who suggested to approximate
the shape of the bump profile by the classical
(Lorentzian) dispersion profile.
After examination of the IUE
extinction curves for many lines of sight,
Fitzpatrick and Massa~\cite{fm07,fm86,fm90} deduced a single analytical
expression with a small number of parameters describing
 extinction in the region
1150\,\AA \, $\leq \lambda <$ 2700\,\AA\,
($x \equiv \lambda^{-1}$)
\begin{equation}
A^{(n)}(x)=\frac{E(\lambda-V)}{E(B-V)}=
\left\{
\begin{array}{ll}
c_1+c_2x+c_3D(x,W,x_0),               & x \leq c_5, \\
c_1+c_2x+c_3D(x,W,x_0)+c_4(x-c_5)^2,  & x > c_5,
\end{array}
          \right.
\label{fm_par}
\end{equation}
where
\begin{equation}
D(x,W,x_0)=\frac{x^2}
{(x^2-x^2_0)^2+x^2W^2}. \label{e2200}
\end{equation}
Originally, the value of $c_5$ was fixed at 5.9\,$\mkm^{-1}$ \cite{fm90}.
Equation~(\ref{fm_par}) consists of: {\it i}) a Lorentzian-like  bump term (requiring three
parameters, corresponding to the bump width $W$, position $x_0$, and
strength $c_3$), {\it ii}) a far-UV curvature term
(two parameters $c_4$ and $c_5$; see Fig.~\ref{fext} for illustration),
and {\it iii}) a linear term underlying the bump and the far-UV (two parameters
$c_1$ and $c_2$).
The parameters of the average extinction curve presented in Fig.~\ref{fext}
are: $R_V=3.001$, $x_0=4.592\,\mkm^{-1}$, $W=0.922\,\mkm^{-1}$,
$c_1=-0.175$, $c_2=0.807$, $c_3=2.991$, $c_4=0.319$, $c_5=6.097$ \cite{fm07}.

\begin{figure}[htb]
\centerline{
\resizebox{10cm}{!}{\includegraphics{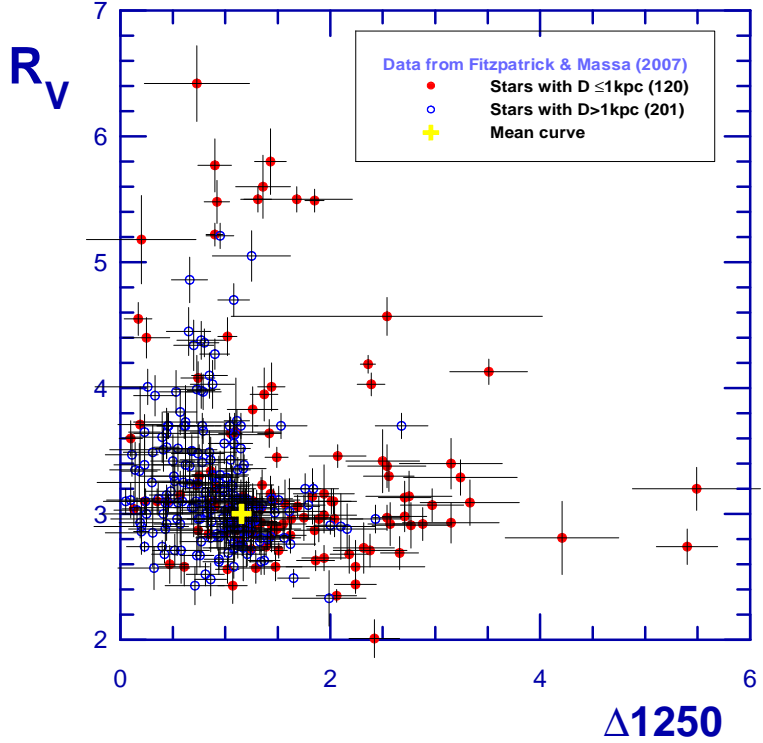}}
}
\caption
{The coefficient $R_V$ in dependence on the quantity
$\Delta 1250=c_4 (8.0-c_5)^2$ showing the strength of the far-UV extinction
curvature in the direction of 321 stars with known distances from \cite{fm07}.
Filled and open circles show data for stars with the distances
$D \leq 1$\,kpc and $D>1$\,kpc, respectively.
Cross corresponds to the average Galactic extinction curve.
}
\label{fuv}
\end{figure}

Fitzpatrick and Massa~\cite{fm07} note that there is no correlation between
the UV and IR portions of the Galactic extinction curves.
This fact is illustrated by Fig.~\ref{fuv} where the dependence of
the coefficient $R_V$ on the quantity describing the strength of
the far-UV extinction curvature $\Delta 1250=c_4 (8.0-c_5)^2$ is shown.
Absence of the correlation between $R_V$ and $\Delta 1250$
contradicts the often used representation of
the extinction curves from the UV to IR as a one-parameter
family dependent on  $R_V$ (so called CCM model introduced by
Cardelli, Clayton and Mathis \cite{ccm88,ccm89}).
As  mentioned in \cite{fm07}, the relations between  $R_V$ and UV extinction
found in \cite{ccm88,ccm89} can arise from sample selection and methodology.
Evidently, some bias relates to the number of clouds available in the
line of sight, i.e. to the distance to the star.
There exists
a wide scatter of the data for nearby stars in comparison with the distant
stars (see Fig.~\ref{fuv}). A large difference in extinction for the nearby
stars observed through single clouds was first noted by Kre{\l}owski
and Wegner \cite{kre89}.
It should be mentioned that the major part of anomalous or peculiar
extinction sightlines studied so far are related to not very distant
stars \cite{mazbar08,mazbar11,zon11,ks12,rai12}.



\subsection{Interpretation: homogeneous spheres}\label{i-hom}

The extinction of stellar radiation at the wavelength $\lambda$
after passing a dust cloud is equal to
\be
A(\lambda)=-2.5\,\log\, {I(\lambda)}/{I_0(\lambda)}\approx
1.086\tau(\lambda),
\ee
where $I_0(\lambda)$ is the source (star) intensity,  $\tau(\lambda)$
the optical thickness which can be found as
the total extinction cross-section of all particles types along the line of
sight in a given direction.
Interpretation of the \is extinction is  often  performed
using homogeneous spherical particles of various size
distributions. Then, the wavelength dependence of extinction
can be calculated as
\begin{equation}
A(\lambda)= 1.086 \, \sum_j \int_0^D \int\limits_{r_{\rm s,\min,j}}^{r_{\rm s,\max,j}}
C_{{\rm ext},j}(m,r_{\rm s},\lambda)\, n_{j}(r_{\rm s})\,{\rm d}r_{{\rm s},j} \,{\rm d}l.
\label{ext}\end{equation}
Here, $n_{j}(r_{\rm s})$ is the size distribution of
spherical dust grains of the type $j$ and radius $r_{\rm s}$ with
the lower cut-off $r_{\rm s,\min}$ and the upper cut-off $r_{\rm s,\max}$,
$C_{\rm ext}(...) = \pi r_{\rm s}^2 \, Q_{\rm ext}(...)$ the extinction
cross-section, $Q_{\rm ext}$ the extinction efficiency factor, $m$ the refractive
index, $D$ the distance to a star. From Eq.~(\ref{ext}), an
important conclusion follows:
{the wavelength dependence of \is extinction is completely
determined by the wavelength dependence of the
extinction efficiency $Q_{\rm ext}$.}

\afterpage{\clearpage}

\begin{figure}[hb]
\bc
\resizebox{10cm}{!}{\includegraphics{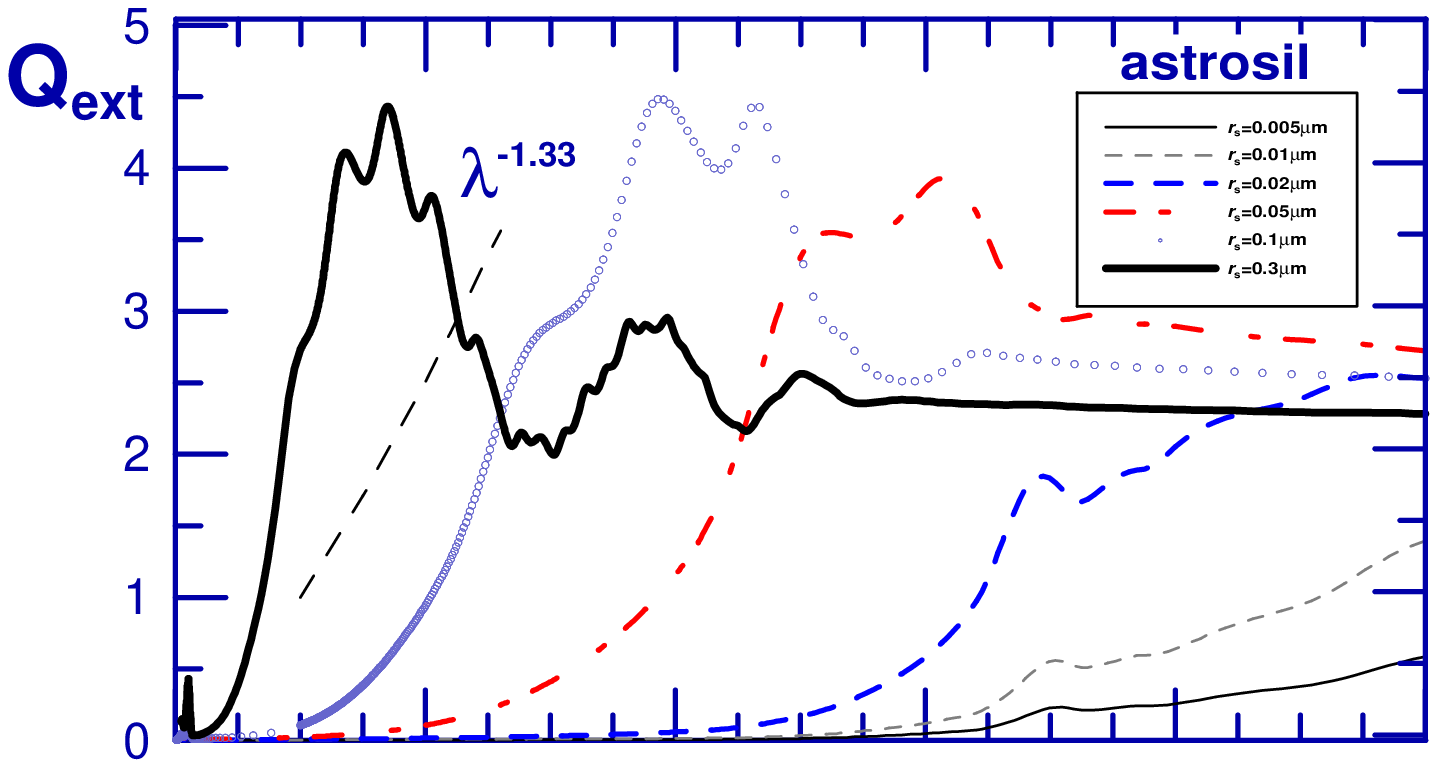}}
\resizebox{10cm}{!}{\includegraphics{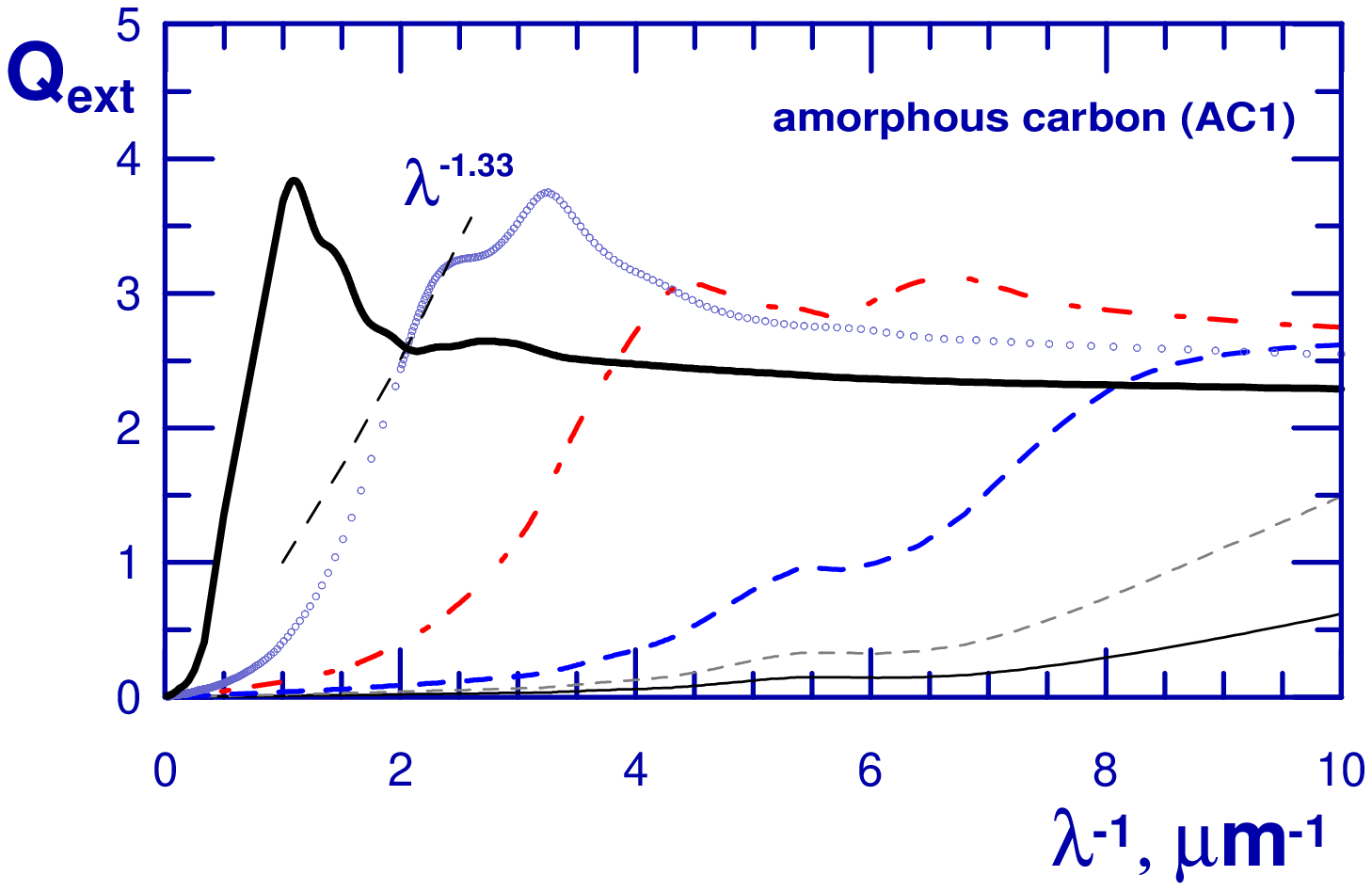}}
\protect\vspace*{-0.4cm}
\caption
{Wavelength dependence of the extinction efficiency factors
for homogeneous spherical particles of different sizes
consisting of astronomical silicate and amorphous carbon.
The dashed segment shows the approximate wavelength dependence of the mean
Galactic extinction curve at optical wavelengths. Adapted from \cite{v04}.}\label{fq_w}
\ec
\end{figure}
\protect\vspace*{-0.8cm}
The average \is extinction curve in the visible-near UV can be approximated
by the power law $A(\lambda) \propto \lambda^{-1.3}$ \cite{fm07}.
Such wavelength dependence can be produced by submicron-sized particles
of the typical radius
$\langle r \rangle \approx 0.05 - 0.1 \, \mkm$.
In this case, for more absorbing materials like amorphous carbon or iron
we have smaller particles and for less absorbing materials like silicate
or ice we need  larger particles (see Fig.~\ref{fq_w}).
So, {from the wavelength dependence of extinction
only the product of the typical particle size and
the refractive index
$\langle r \rangle  \, |m-1| \approx {\rm const.}$
can be determined,
but not the size or chemical composition
of dust grains separately}.
In order to solve this problem,
the dust-phase abundances
of the main elements forming dust (C, O, Mg, Si and Fe)
need to be taken into account
and to reproduce the absolute extinction. Unfortunately,
despite numerous observations of the interstellar absorption lines
(see a compilation of Gudennavar et al. \cite{guden12})
abundances with good accuracy are
known just for a restricted number of diffuse and translucent
clouds  \cite{j09,vh10,parv12}.

\begin{figure}[htb]
\centerline{
\resizebox{7.0cm}{!}{\includegraphics{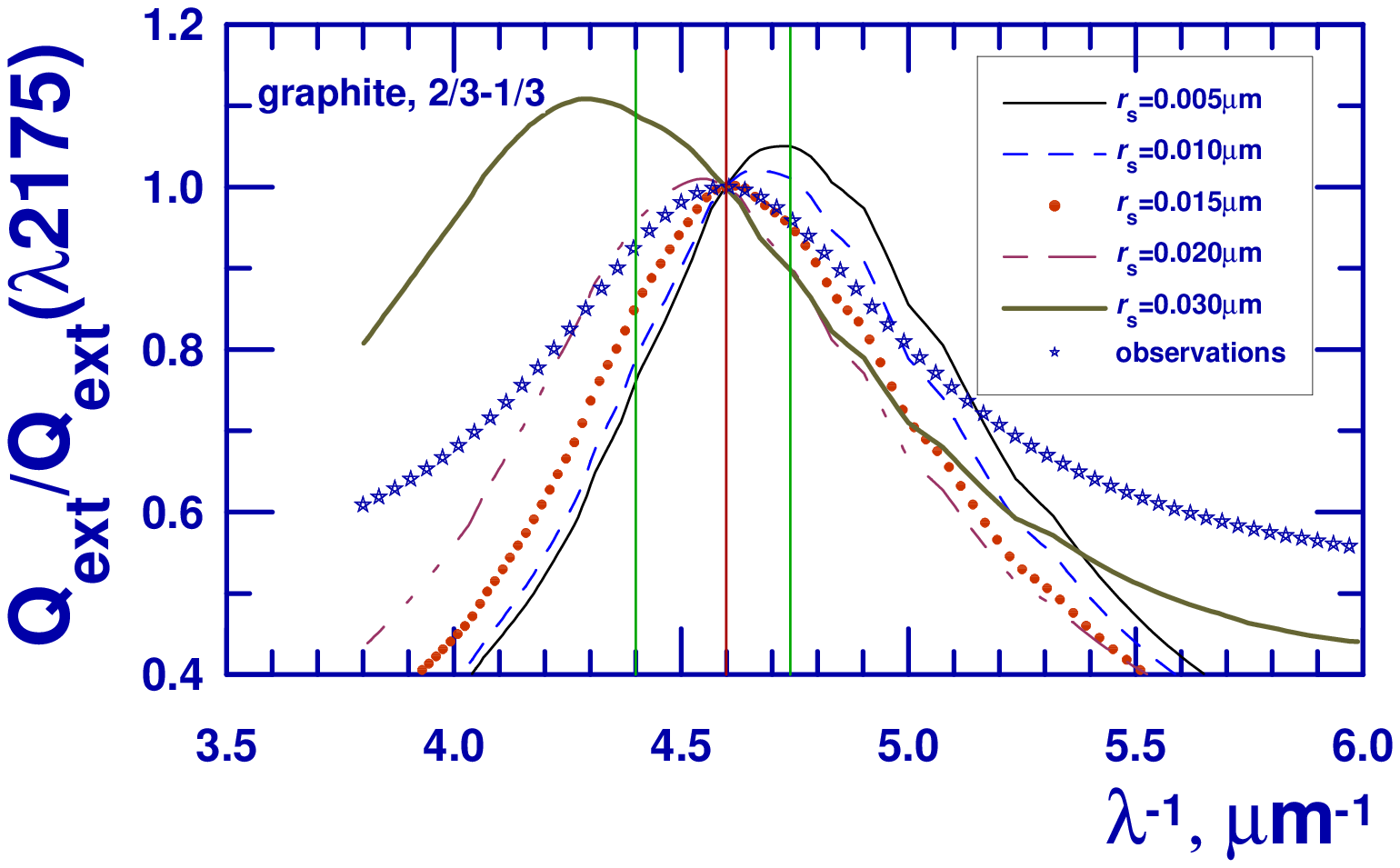}}
\resizebox{7.0cm}{!}{\includegraphics{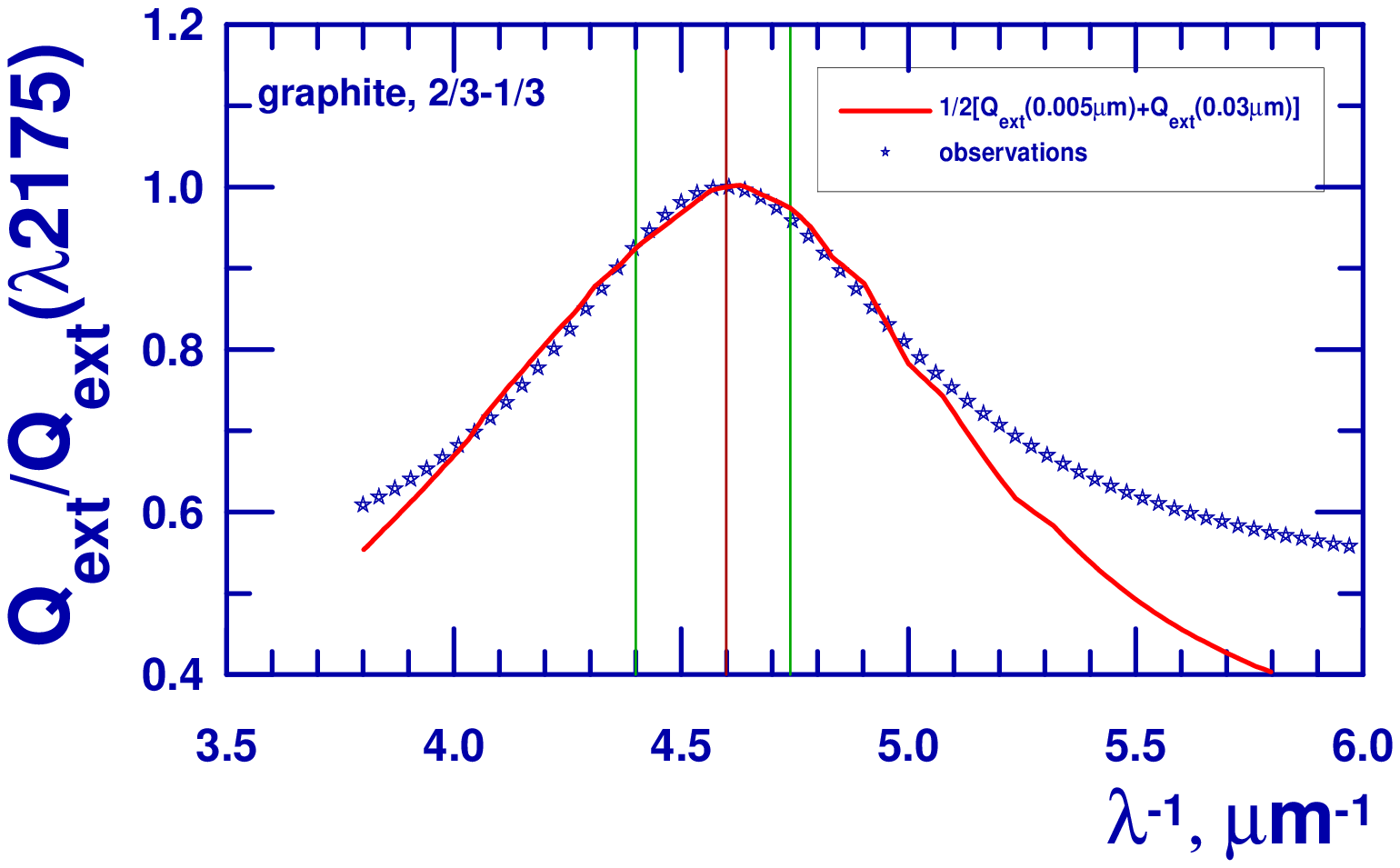}}
}
\caption
{Normalized extinction efficiencies for graphite spheres.
The curve marked as ``observations'' corresponds to the wavelength
dependence of the UV bump  given by the mean Galactic extinction curve.
The central position of the observed UV bump and its range of variations
are marked.
The left panel shows extinction of single size graphite spheres.
The right panel shows the summary extinction of two graphite spheres
with radii $r_{\rm s} = 0.005 \,\mkm$ and
$r_{\rm s} = 0.03 \,\mkm$ (from left panel) taken in equal proportions.
All calculations were made in the ``2/3--1/3'' approximation
for the averaged extinction factors
$
Q_{\rm ext} = {2}/{3}\, Q_{\rm ext}(\varepsilon_{\bot}) +
              {1}/{3}\, Q_{\rm ext}(\varepsilon_{||}),
$
where $\varepsilon_{\bot}$ and $\varepsilon_{||}$ are the dielectric
functions for two cases of orientation of the electric field relative
to the basal plane of graphite.
Adapted from \cite{v04}.
}\label{fgran}
\end{figure}

Another  problem having many solutions is identification of
the UV bump near $\lambda$\,2175\,\AA.
Various materials with isotropic and anisotropic properties such as
silicate (enstatite), irradiated quartz, oxides (MgO, CaO), organic molecules
have been considered as carrier candidates
(see discussion in \cite{w03,v04}).
However, the position and width of the bump are strongly suggestive
of $\pi \rightarrow \pi^*$ transitions in graphitic or aromatic
carbonaceous species dominating by sp$^2$ bonding.
Therefore, small graphite particles and polycyclic aromatic hydrocarbons
(PAH molecules) are considered as the favourite materials \cite{wd01,ld01,cc08}.
Unfortunately, the reliable identification of the carrier remains unknown.
The attempts to find the UV or visual bands of PAHs have
failed  \cite{cla03,gre11}. We cannot even determine the size of
graphite spheres responsible for the $\lambda$~2175~\AA\, feature
(see Fig.~\ref{fgran}).
The profiles with the central position
near $\lambda_0^{-1} = 4.6 \,\mkm^{-1}$ can be obtained if we take
particles with the radius $r_{\rm s} \approx 0.015\,\mkm$
(Fig.~\ref{fgran}, left panel). Although for single-size particles
the width of the calculated profiles is smaller than
the observed one, a simple bi-modal size
distribution allows a fit to both the position and the width of
the mean Galactic profile (Fig.~\ref{fgran}, right panel).


The far-UV extinction 
can be explained by tiny particles of the typical radius
$\langle r \rangle \approx 0.01 - 0.03 \, \,\mkm$ (see Fig.~\ref{fq_w}).
The number density of such grains is $\sim 1000$ times large  than
the submicron particles producing the visual-near-IR extinction
\cite{gre78}. Because of temperature fluctuations, such particles
are protected from growth by accretion in the interstellar clouds.
The far-UV rise of extinction may be also fitted as
the low-energy side of $\sigma \rightarrow \sigma^*$ transitions in PAHs
(see \cite{cc10} for discussion).


By using particles of different chemical composition and applying Eq.~(\ref{ext})
it is possible to interpret the interstellar extinction and to reconstruct
the dust size distribution.
In the pioneer work of Oort and van de Hulst~\cite{ovdh46}
the size distribution of icy grains was found in the tabular form.
Later, Greenberg~\cite{gre68} fitted it with an exponential function.
By using minimization of the $\chi^2$ statistic,
Mathis et al. \cite{mrn} reconstructed the power-law size distribution
for silicate and graphite particles.
These two simplest size distributions contain the only parameter
(except for the lower and upper cut-offs, see Table~\ref{t-size}).
More complicated two-parameter distributions were applied
by Wickramasinghe and Guillaume \cite{wg65} and
Wickramasinghe and Nandy \cite{wn71} in order to fit the mean
extinction curve with graphite grains and a mixture of
graphite, iron and silicate grains, respectively.
A comprehensive discussion of the early attempts to model the interstellar
extinction can be found in the review of Wickramasinghe and Nandy \cite{wn72}.

The size distributions of dust grains can be also found
from extinction measurements by solving the inverse problem.
It has been made by the maximum entropy method \cite{kmh94} and
by Tikhonov's method of regularization \cite{zda04,zkw96,zkw98}.
The obtained size distributions were approximated by
rather complicated functions containing up to 14 parameters
(see Table~\ref{t-size}).
\begin{table}[htb]
\bc
\caption{Dust size distributions used for interpretation of \is extinction.}\label{t-size}
\begin{tabular}{lc}
\noalign{\smallskip}\hline\noalign{\smallskip}
Author(s) (year) reference; size distribution function & $N_{\rm parameters}$ \\
\noalign{\smallskip}\hline\noalign{\smallskip}
Greenberg (1968) \cite{gre68}; exponential &\\
$\displaystyle n(r_{\rm s}) \propto \exp \left[-5 \left( {r_{\rm s}}/{r_{\rm s0}} \right )^3\right]$ & 1  \\
\noalign{\smallskip}
Isobe (1973) \cite{iso73}; exponential &\\
$\displaystyle n(r_{\rm s}) \propto \exp \left[- \left(r_{\rm s}/r_{\rm s0} \right )\right]$ & 1  \\
\noalign{\smallskip}
Mathis et al. (1977) \cite{mrn}; power-law; MRN mixture & \\
 $\displaystyle n(r_{\rm s}) \propto r_{\rm s}^{-q}$ & 1  \\
\noalign{\smallskip}
Wickramasinghe and Guillaume (1965) \cite{wg65}; normal &\\
$\displaystyle n(r_{\rm s}) \propto \exp \left[- \left(r_{\rm s}- \bar r_{\rm s} \right )^2/(2 \sigma^2) \right]$ & 2  \\
\noalign{\smallskip}
Wickramasinghe and Nandy  (1971) \cite{wn71}; lognormal  &\\
$\displaystyle n(r_{\rm s}) \propto r_{\rm s}^{r_{\rm 1}} \exp \left[-1/2 \left(r_{\rm s}/r_{\rm 2} \right )^3 \right]$ & 2  \\
\noalign{\smallskip}
Kim et al. (1994)  \cite{kmh94}; power-law with exponential decay &\\
 $\displaystyle n(r_{\rm s}) \propto r_{\rm s}^{- \gamma} \, \exp (-r_{\rm s}/r_{\rm sb})$ & 2  \\
\noalign{\smallskip}
Mathis (1996) \cite{m96}; power-law with exponential decay &\\
 $\displaystyle n(r_{\rm s}) \propto r_{\rm s}^{- \gamma_0} \, \exp [-(\gamma_1r_{\rm s} + \gamma_2/r_{\rm s} + \gamma_3/r_{\rm s}^2)] $ & 4  \\
\noalign{\smallskip}
Weingartner and Draine (2001) \cite{wd01}; two lognormal &\\
$\displaystyle n(r_{\rm s}) \propto {\cal D}_{\rm C}(r_{\rm s}) + \frac{{\cal C}_{\rm C,\,Si}}{r_{\rm s}}
\left(\frac{r_{\rm s}}{r_{\rm t;\,C,\,Si}}\right)^{\alpha_{\rm C,\,Si}} $&\\
 $\times \left\{
\begin{array}{ll}
1 + \beta_{\rm C,\,Si}r_{\rm s}/r_{\rm t;\,C,\,Si},  & \beta \geq 0 \\
(1 - \beta_{\rm C,\,Si}r_{\rm s}/r_{\rm t;\,C,\,Si})^{-1}, &  \beta < 0
\end{array}  \right\}$ &\\
 $\times \left\{
\begin{array}{ll}
1,  & 3.5 \,{\rm \AA} < r_{\rm s} < r_{\rm t;\,C,\,Si} \\
\exp \{- [(r_{\rm s} - r_{\rm t;\,C,\,Si})/r_{\rm t;\,C,\,Si}]^3\},  &  r_{\rm s} > r_{\rm t;\,C,\,Si}
\end{array} \right\} $ & 11   \\
\noalign{\smallskip}
Zubko et al. (2004) \cite{zda04};  &\\
$\displaystyle \log n(r_{\rm s}) = c_0 + b_0\log(r_{\rm s}) - b_1|\log(r_{\rm s}/a_1)|^{m1}$&\\
\noalign{\smallskip}
$\displaystyle - b_2|\log(r_{\rm s}/a_2)|^{m2} -b_3|r_{\rm s}-a_3|^{m3}
-b_4|r_{\rm s}-a_4|^{m4}$& 14\\
\noalign{\smallskip}\hline
\end{tabular}
\ec
\end{table}
Complexity of the size distributions used in \cite{wd01,zda04}
is explained by the challenge in reproducing
the diffuse Galactic IR emission as well.

\clearpage

\subsection{Interpretation: inhomogeneous and composite particles}

Progress in observations, the light scattering and grain growth
theories gave rise to new dust models with grains more complicated
than homogeneous spheres.

Wickramasinghe~\cite{w63,w70} was the first who studied the optical
properties of core-mantle grains which could grow in interstellar clouds due
to accretion of volatile elements on refractory particles.
He calculated extinction produced by graphite core--ice mantle and
silicate core--ice mantle spheres. Extensive calculations of
extinction for graphite core--ice mantle particles were also made by
Greenberg \cite{gre68} who later proposed the existence of
particles with silicate cores
coated by a layer of organic material in diffuse clouds and
silicate-organic-ice grains in molecular clouds \cite{gre78}.
Such grains were  a component of the dust mixture reproducing
interstellar extinction  \cite{gl96}.



The growth of interstellar grains due to their coagulation in dense
molecular cloud cores may result in formation of grain aggregates with
large voids \cite{dh95}. The internal structure of such composite
grains can be very complicated, and their optical properties cannot
be described by the model of core--mantle spheres.
Exact calculations are possible for complex aggregates
of rather small sizes  \cite{fi06,bds2,min09}. Therefore, very complicated
particles are replaced by more simple ``optically equivalent'' ones.
A very popular approach is to make calculations using the Mie theory
for homogeneous spheres with an average refractive index derived from
one of the mixing rules of the
effective medium theory (EMT; see, e.g., \cite{m96,zda04,mw89} and
Table~\ref{t-mod}).
\begin{table}[htb]
\bc
\caption{Models of inhomogeneous spherical grains used for interpretation of \is extinction.}\label{t-mod}
\begin{tabular}{ll}
\noalign{\smallskip}\hline\noalign{\smallskip}
Author(s) (year) reference & Model \\
\noalign{\smallskip}\hline\noalign{\smallskip}
Wickramasinghe (1963) \cite{w63};  & graphite core--ice mantle \\
~Greenberg (1968) \cite{gre68} &  \\
\noalign{\smallskip}
Wickramasinghe  (1970)\cite{w70} & silicate core--ice mantle \\
\noalign{\smallskip}
Greenberg, Li  (1996) \cite{gl96} & silicate core--organic mantle \\
\noalign{\smallskip}
Mathis and  Whiffen (1989) \cite{mw89};  & EMT-Mie: silicate  + amorphous   \\
 ~Mathis (1996) \cite{m96}         & carbon + iron + voids  \\
\noalign{\smallskip}
Zubko et al. (1998, 2004) \cite{zda04,zkw98} & EMT-Mie: silicate + organic  \\
                          & refractory + water ice + voids \\
\noalign{\smallskip}
Vaidya  et al. (2001) \cite{va01} & silicates with graphite inclusions \\
\noalign{\smallskip}
Voshchinnikov and Mathis (1999) \cite{vm99}; & multi-layered: vacuum/silicate/  \\
~Voshchinnikov  et al. (2006) \cite{vihd06}    & amorpous carbon   \\
\noalign{\smallskip}
Iat\`\i \,et al. (2008) \cite{iatietal08}; & four-layered: vacuum-- \\
~Cecchi-Pestellini et al. (2010) \cite{cc10};  &  silicate--sp$^2$-carbon-- \\
~Zonca et al. (2011) \cite{zon11} &  sp$^3$-carbon \\
\noalign{\smallskip}
Rai and Rastogi (2012) \cite{rai12} & nanodiamonds coated by \\
   &   amorpous carbon  or  graphite \\
\noalign{\smallskip}\hline
\end{tabular}
\ec
\end{table}
Another possibility to treat composite aggregate grains is
to consider multi-layered particles. As
shown by Voshchinnikov and Mathis \cite{vm99} for spheres and
by Farafonov and Voshchinnikov \cite{fv12} for spheroids,
the scattering characteristics of layered particles slightly depend on
the order of materials and become close
to some ``average'' ones, when the number of layers exceeds $15 - 20$.
According to estimates made in \cite{vih05}, the optical properties of layered
particles resemble those of heterogeneous particles having inclusions
of various sizes while the EMT-Mie approach can be used if the particles
have small (in comparison with the wavelength of the incident radiation)
``Rayleigh'' inclusions.

\begin{figure}[htb]
\bc
\resizebox{12cm}{!}{\includegraphics{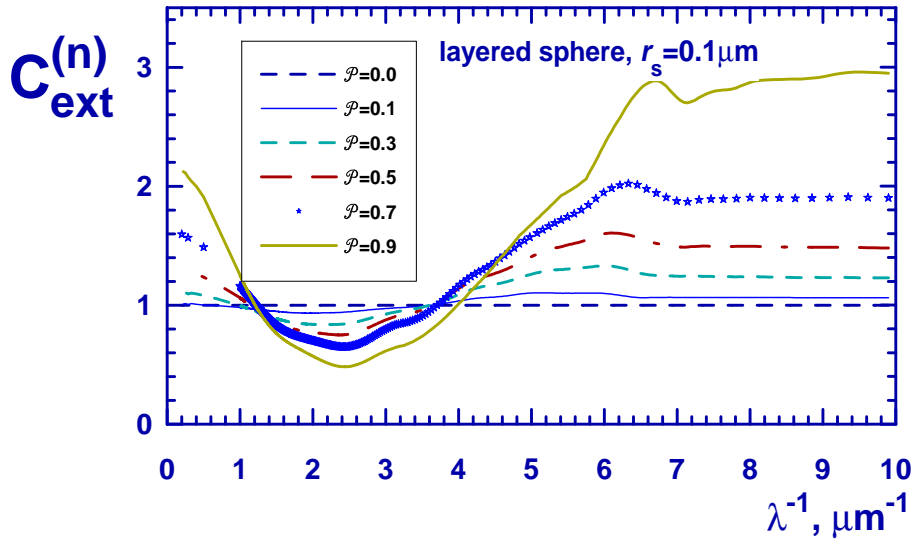}}
\caption{Wavelength dependence of the normalized extinction cross section
(see Eq.~(\ref{cn}))
for multi-layered spherical particles with $r_{\rm s,\,compact}=0.1\,\mkm$.
The particles are of the same mass but of different porosity.
If $C_{\rm ext}^{\rm (n)} >1$, extinction of a porous particle is larger
than that of a compact one of the same mass.
Adapted from \cite{vihd06}.
}\label{pp01}\ec
\end{figure}

Inhomogeneous and composite particles have an advantage over homogeneous
ones as there exists the possibility of including vacuum as one
of the materials.
The new dust models with fluffy, porous particles are able to produce the same
extinction with a smaller amount of solid material than
dust models with compact particles.
The amount of vacuum in a particle can be characterized by its porosity
${\cal P}$ ($0 \leq {\cal P} < 1$)
\be
{\cal P} = V_{\rm vac} /V_{\rm total}
= 1 - V_{\rm solid} /V_{\rm total}. \label{por}
\ee
The role of porosity in extinction is seen from Fig.~\ref{pp01} that gives
the wavelength dependence of the normalized cross section
\bea
C_{\rm ext}^{\rm (n)} = \frac{C_{\rm ext}({\rm porous \, grain})}
{C_{\rm ext}({\rm  compact \, grain \, of \, same \, mass})} = \nonumber \\
\,\,\,\,\,\,\, (1-{\cal P})^{-2/3}\,  \frac{Q_{\rm ext}({\rm porous \, grain})}
{Q_{\rm ext}({\rm  compact \, grain \, of \, same \, mass})}. \label{cn}
\eea
This quantity shows how porosity influences the extinction cross section.
As follows from Fig.~\ref{pp01}, as ${\cal P}$ increases
the model predicts a growth of
extinction of porous particles in the far-UV and a decrease in the
visual--near-UV.
In comparison with compact grains, layered particles can also produce rather
large extinction in the near-IR.
This is especially important for the explanation of the flat extinction
across the $3 - 8\,\mkm$ wavelength range measured for several
lines of sight (see \cite{fritz11,vihd06} for discussion).
It is also seen from Fig.~\ref{pp01} that an addition of vacuum into
particles does not
lead to a  growth of extinction at all wavelengths and
material saving\footnote{This supports a conclusion
of Li \cite{li05b} that an interpretation of the observed interstellar
extinction curve using only very porous grains should
not give any gain in dust-phase abundances.}.
Evidently, the final conclusion can be made after
detailed comparison of the observations with theoretical calculations
at many wavelengths.

Table~\ref{t-mod} contains information about models of inhomogeneous
spherical grains used for the interpretation of \is extinction.
Inhomogeneous non-spherical
particles of the simplest shapes (cylinders, spheroids)
are also used for simultaneous interpretation of \is extinction
and polarization (see Table~\ref{p-mod}). In this case
major attention has been paid to the modelling of polarization
because extinction usually has only a slight
dependence on the particle shape and orientation \cite{v04,vd08}.

The models discussed in this Section
were first applied to interpretating
the average Galactic extinction curve.
Modelling of extinction in several  sightlines
has been also performed
\cite{mazbar08,mazbar11,rai12,wd01,zkw96,zkw98,vihd06,chkru83,aa95,lg98,laretal00}.
The models include multi-component mixtures of bare particles with
a rather complicated size distribution function or/and inhomogeneous
particles and PAHs and often took into account  abundance
restrictions. Especially popular is the direction to the halo
star HD~210121 with very high UV extinction.
The extinction for this star has been modeled by
Li and Greenberg~\cite{lg98} with a mixture of silicate core--organic mantle
spheres,  bare spheres and PAHs,
by Larson et al. \cite{laretal00} and by  Clayton et al. \cite{clayal03}
with two-component (silicate, graphite)
and three-component (silicate, graphite, amorphous carbon) models,
by Weingartner and Draine~\cite{wd01} who considered
the mixture of carbonaceous and silicate grains with the size
distribution given in Table~\ref{t-size},
and by Rai and Rastogi \cite{rai12} who used a silicate-graphite mixture
and nanodiamonds coated by carbon. This list illustrates the non-uniquiness
of parameters of dust grains  obtained from modelling.

A similar conclusion about the ambiguity of the modelling
follows from several attempts to interpret the peculiar extinction
curves characterized by a broad $\lambda$~2175\,\AA\, bump and a steep far-UV
rise or a sharp $\lambda$~2175\,\AA\, bump and flat far-UV
extinction (see Fig.~\ref{fext}).
Using the model of Weingartner and Draine~\cite{wd01} (see Table~\ref{t-size}),
Mazzei and Barbaro \cite{mazbar11} derived the parameters of the size
distribution for 64 stars. Variations of the parameters were attributed to the
selective grain destruction in both shocks and grain-grain collisions.
Zonca et al. \cite{zon11} found an excellent fit for different
extinction curves for 15 sightlines with the mixture of layered porous
grains for reproduction of the near-IR-visual extinction and PAHs
to account for the bump and far-UV extinction.
Large grains consisting of silicates coated by layers from graphitic
and polymeric amorphous carbons (see Table~\ref{t-mod})
were suggested in the model of Jones et al. \cite{jdw90}
(see \cite{iati12} for more details).
Rai and Rastogi \cite{rai12} analyzed anomalous extinction curves in the
direction of 10 stars and showed that a very good match with the far-UV
rise of extinction was obtained if to include nanodiamonds coated by
graphite or amorphous carbon as a component of the silicate-graphite mixture.

Summarizing this discussion of the \is extinction, it should be noted that
there is
a wide diversity in the models and the non-uniqueness in the results.
In spite of numerous attempts to use very complicated
inhomogeneous particles, the Mie theory for homogeneous spheres
keeps its leading position as a main tool for the interpretation
of \is extinction. Further progress in the investigations should
include a clear role for PAHs and modelling of the extinction
on the basis of interstellar abundances in selected directions.
A sofisticated undertstanding of the origin of UV extinction,
the available solid-state material and the grain growth process
would stimilate going from simple Mie theory to
justified  models of complex particles.


\section{Polarization}

\subsection{Observations: Serkowski curve and polarizing efficiency}\label{p-obs}

Interstellar linear polarization is caused by the linear dichroism of
the interstellar medium due to the presence of non-spherical aligned
grains. Dust grains must have sizes close to the wavelength
of the incident radiation and specific magnetic properties  to efficiently
interact with the interstellar magnetic field.
The direction of alignment must not coincide with the line of sight and
there must be no cancellation of polarization during the
propagation of radiation through the interstellar medium.

Interstellar polarization was discovered in 1949 by Hiltner~\cite{hi49},
Hall \cite{hall49} and Dombrovskii~\cite{d49} in the course of the search
for the Sobolev-Chan\-dra\-se\-khar effect\footnote{Sobolev and Chandrasekhar have
shown that the polarization
of radiation at the limb of a star due to the electronic (Thomson) scattering
should reach $\sim 12\%$. Eclipsing binaries with extended atmospheres
have been suggested in order to observe the effect.}.
The wavelength dependence of polarization $P(\lambda)$ in the visible part of
spectrum is described by an empirical formula suggested by Serkowski~\cite{serk73}
\be
P(\lambda)/P_{\max} = \exp [-K \ln^2 (\lambda_{\max}/\lambda)].
\label{serkk}
\ee
This formula has three parameters: the maximum degree of polarization $P_{\max}$,
the wavelength corresponding to it $\lambda_{\max}$ and
the coefficient $K$ characterizing the width of the Serkowski curve.
Initially, the coefficient $K$ was chosen by Serkowski~\cite{serk73,smf75}
to be equal to 1.15$^($\footnote{The Serkowski curve
is just one of possible approximations of the observed dependence
$P(\lambda)$. For example, Wolstencroft and Smith \cite{ws84} have suggested the
representation
$$
P(\lambda)/P_{\max} = 2^{\cal K} (\lambda/\lambda_{\max} +
\lambda_{\max}/\lambda)^{-{\cal K}}.
$$
When ${\cal K}=2.25$ this curve lies within 1\,\% of the Serkowski curve
with $K=1.15$ in the wavelength interval $0.22 - 1.40\,\,\mkm$. \\
Another approximation has been proposed by Efimov \cite{ef09}
$$
P(\lambda)/P_{\max} = [(\lambda/\lambda_{\max})\, \exp(1-
\lambda/\lambda_{\max})]^{\beta},
$$
where the index $\beta$ is proportional to $K$.}$^)$.

\begin{figure}[htb]
\centerline{\resizebox{9.3cm}{!}{\includegraphics{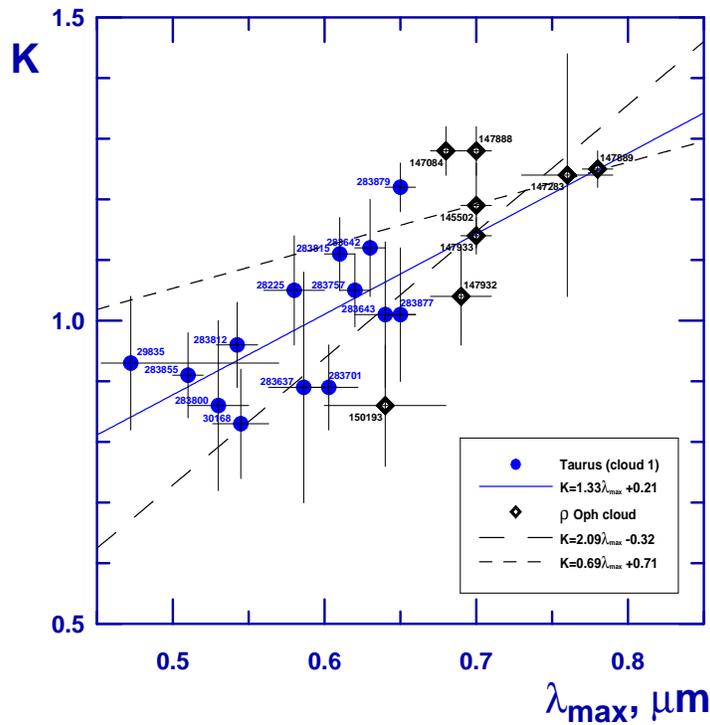}}}
\caption
{The coefficients $K$ of the Serkowski curve (\ref{serkk})
in dependence on
the wavelength of the maximum polarization $\lambda_{\max}$.
The data were taken from \cite{ef09,wetal01} for the Taurus dark cloud
and from \cite{maretal92} for $\rho$~Oph cloud.
The HD numbers of stars are marked.
Taurus cloud 1 contains 14 stars with  similar positional angles
of polarization $\theta_{\rm gal.}=145\degr - 175\degr$ (see \cite{vy12}
for details).
HD~150193 is Herbig Ae/Be star with intrinsic polarization \cite{rod08}.
The linear fits  for $\rho$~Oph cloud
are shown for cases with and without HD~150193.
}
\label{klm}
\end{figure}

The values of $P_{\max}$ in the diffuse interstellar medium usually
do not exceed 10\%.
The ratios $P_{\max}/E({B-V})$ and $P_{\max}/A_V$ determine the
polarizing efficiency of the \is medium in a selected direction.
There exist empirically found upper limits on these ratios \cite{smf75}
\be
P_{\max}/E({B-V}) \la 9 \%/{\rm mag}
\,\,\,\,\,\,\,\,\,{\rm and}\,\,\,\,\,\,\,\,\,
P_{\max}/A_V \la 3 \%/{\rm mag}\,.
\label{pebv}\ee
The mean value of $\lambda_{\max}$ is 0.55\,$\mkm$ although there are
directions for which $\lambda_{\max}$ is smaller than 0.4\,$\mkm$
or larger than 0.8\,$\mkm$ \cite{ef09} (see also Fig.~\ref{klm}).

Using observations of about 50 southern stars
Whittet and van Breda \cite{wbreda78} established a relation between
the parameters of the extinction and polarization curves
$R_{V} = (5.6 \pm 0.3) \, \lambda_{\max},$ where $\lambda_{\max}$ is in $\mkm$.
However, further investigations of separate clouds
questioned this correlation (e.g., \cite{chkru83,wetal01,ap07}).


The connection between the coefficient $K$ and the width of the
normalized curve of interstellar linear polarization $W$
is given by the relation
$$
W = \exp [(\ln 2/K)^{1/2}] -  \exp [-(\ln 2/K)^{1/2}].
\label{wk}
$$
%
Treating $K$ as a third free parameter of the
Serkowski curve, Whittet  et al.~\cite{wetal92} evaluated the dependence
between $K$ and $\lambda_{\max}$ on the basis of observations for 109 stars
\be
K = (1.66 \pm 0.09) \lambda_{\max} + (0.01 \pm 0.05).
\label{k92}
\ee
The coefficients of the linear function (\ref{k92})
for different regions may strongly
deviate from the average values (see Fig.~\ref{klm} where the data
for the Taurus dark cloud and the $\rho$~Oph cloud are plotted).

In parallel with the  positive correlation between $K$ and $\lambda_{\max}$,
the negative correlation between the polarization efficiency
$P_{\max}/A_V$ and $\lambda_{\max}$ for stars in separate \is clouds
and associations is observed
\cite{wetal01,wetal94,vy10} (see also discussion in Sect.~\ref{tdc}).


The IR continuum polarization for $\lambda > 2.5 \,\mkm$ cannot be represented
by the Serkowski curve with three parameters.
The polarization seems to have a common, universal functional form
independent of the value of  $\lambda_{\max}$ and
its wavelength dependence is given by a power law
$P(\lambda) \propto \lambda^{-(1.6-2.0)}$ \cite{marwh90,maretal92}.
The UV polarization for 28 lines of sight in the Galaxy has been analyzed by
Martin et al.~\cite{mcw99} and  fitted by a Serkowski-like curve.

As interstellar extinction and interstellar polarization
have different wavelength dependencies,
\begin{figure}[ht]
\centerline{
\resizebox{12cm}{!}{\includegraphics{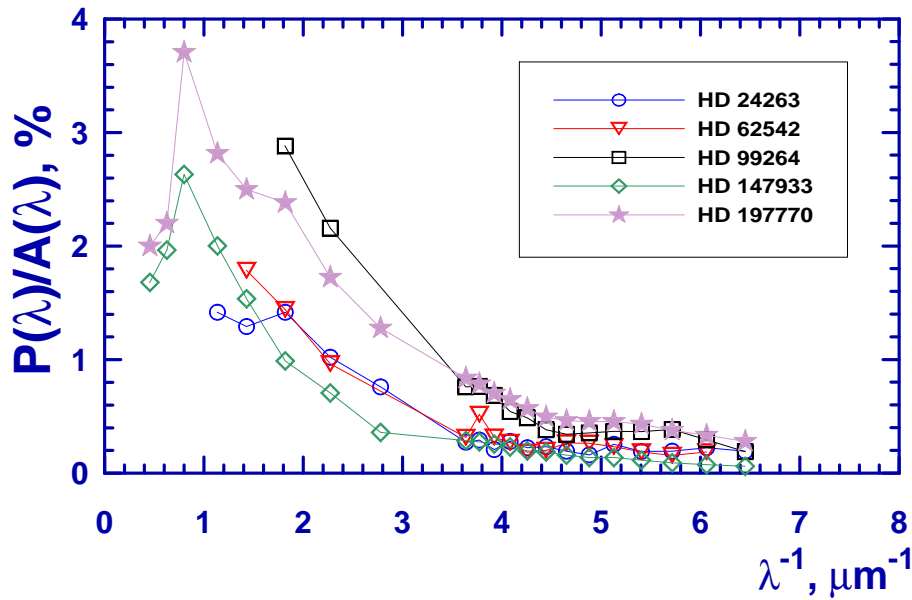}}}
\caption
{The polarizing  efficiency of the interstellar medium
in the direction of five stars.
Observational data were taken from \cite{val04} (extinction) and
\cite{wilk80,andal96} (polarization).
Adapted from \cite{vd08}.}
\label{fig2}
\end{figure}
the polarizing efficiency $P(\lambda)/A(\lambda)$ has a maximum in
the near-IR.
Note that the polarizing  efficiency generally increases with
wavelength for $\lambda \la 1\,\,\mkm$. It may be approximated by
the power-law dependence $P/A \propto \lambda^{\epsilon}$.
For stars presented in Fig.~\ref{fig2} the values of
$\epsilon$ vary from 1.41 for HD~197770 to 2.06 for HD~99264.

Variations of the polarizing efficiency in cold dark clouds and star-forming
regions are of special interest. It was found that in several dark clouds
the rise of polarization with growing extinction was stopped  at
some value of $A_V$ \cite{goodetal95,arceetal98} and that
the polarizing efficiency $P_{\max}/A_V$ or $P_K/A_K$
declines rapidly with increasing extinction
\cite{wetal01,ap07,gwl95,w05,wetal08}.
This fact is usually considered as the evidence for the lower
efficiency of grain alignment in dark clouds in comparison
with diffuse clouds \cite{lgm97}.
But it is possible to propound many other factors influencing
the polarization degree. They are  discussed  qualitatively by
Goodman et al. \cite{goodetal95} (see their Table~4).
Netherveless, there exists the direct evidence for grain alignment in
cold dense environments.
For example, Hough et al. \cite{hou88,hou08},
using spectropolarimetry of the 3.1\,$\,\mkm$ and 4.67\,$\,\mkm$
solid H$_2$O and CO features along the line of sight to Elias 16, a field star
background to the Taurus dark cloud, found that the features were polarized.
This indicates the presence of multi-layered nonspherical grains in
molecular clouds  because solid CO survive at 
$T< 20$\,K and solid H$_2$O at higher temperatures.

An error often made is ignoring the foreground polarization.
This can be well illustrated by
the observational data of Messenger et al. \cite{mess97} who
analysed the interstellar polarization in the Taurus dark cloud.
The authors applied a two-component model and excluded the foreground
polarization for star HD 29647.
\begin{table}[htb]
\bc
\caption{Polarization efficiency for stars in Taurus dark
cloud$^*$.}\label{p-mess}
\begin{tabular}{lcccl}
\noalign{\smallskip}\hline\noalign{\smallskip}
Star    & $P_{\max},\%$  & $A_V$  &$P_{\max}/A_V,\%$ & Comment \\
\noalign{\smallskip}\hline\noalign{\smallskip}
HD 283812 & 6.30 &  1.91  &  3.30 & cloud 1 (foreground) \\
HD 29647  & 2.30 &  3.32  &  0.69 & clouds 1 + 2 \\ 
HD 29647  & 6.17 &  1.41  &  4.38 & only cloud 2  (background) \\
\noalign{\smallskip}\hline
\end{tabular} \ec
{\small $^*$ data from Messenger et al. \cite{mess97}}
\end{table}
Using the data from Tables 1 and 3 of Messenger et al. \cite{mess97}
it is possible to estimate the polarizing efficiency in the foreground
and background clouds.
As follows from Table~\ref{p-mess}, the polarization efficiency in
cloud 2  (background) is very high (cf. (Eq.~\ref{pebv})).
Therefore, interpretation of the observed polarization
instead of the true polarization in the background cloud
is a large mistake for HD 29647.

\subsection{Interpretation: particles and alignment}

\begin{figure}[htb]
\centerline{
\resizebox{8cm}{!}{\includegraphics{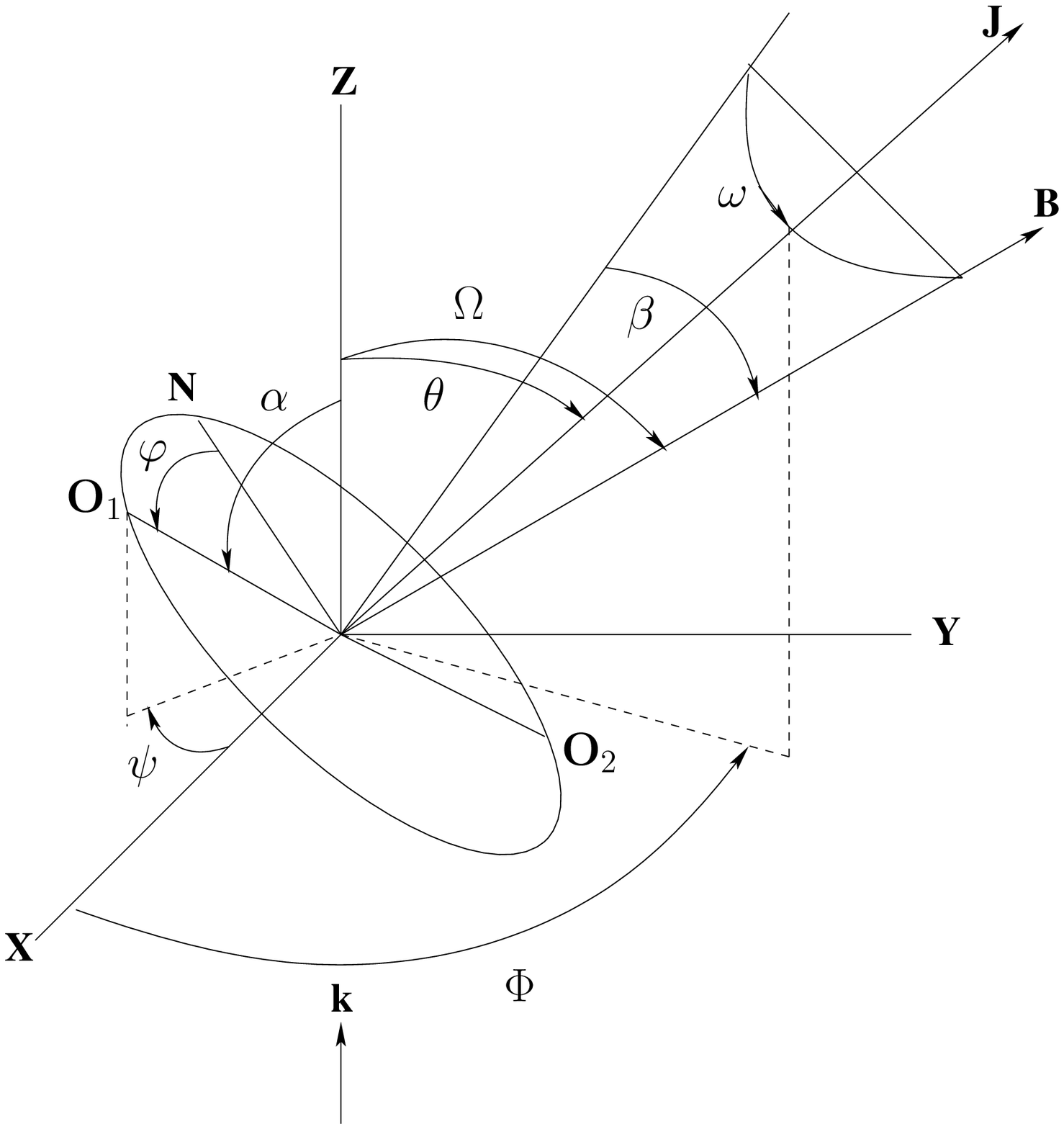}}}
 \caption{Geometrical configuration of a spinning and
 wobbling prolate spheroidal grain.
 The major (symmetry) axis of the particle 
 O$_1$O$_2$ is situated in the spinning plane NO$_1$O$_2$
 which is perpendicular to the angular momentum $\vec{J}$.
 The direction of light propagation $\vec{k}$ is parallel to the $Z$-axis
 and makes the angle $\alpha$ with the particle symmetry axis.
The angle between the line of sight and the magnetic
field is $\Omega$.
 After \cite{dvi10}.
 }
   \label{f-sky}
\end{figure}

The interpretation of polarimetric observations
includes computations of the polarization cross sections
and their averaging over given particles size  and orientation distributions.
The linear polarization of non-polarized stellar radiation passing
through a cloud with a homogeneous magnetic field and rotating  particles
can be found as (cf. Eq.~(\ref{ext}))
\begin{equation}
P(\lambda)= \sum_j \int_0^D \int\limits_{r_{V,\min,j}}^{r_{V,\max,j}}
\overline{C}_{{\rm pol},j}(m,r_{V},\lambda)\, n_{j}(r_{V})\,{\rm d}r_{V,j} \,{\rm d}l
\cdot 100\,\%\,,
\label{pol}\end{equation}
\be
 \overline{C}_{{\rm pol},j}({\lambda})  =  {\frac{2}{\pi^2}}
{\int_{0}^{\pi/2}}{\int_{0}^{\pi}}{\int_{0}^{\pi/2}}
\frac{1}{2} (C^{\rm TM}_{{\rm ext},j}-C^{\rm TE}_{{\rm ext},j}) \,
f_j(\xi, \beta) \, \cos 2{\psi} \, d{\varphi} d{\omega} d{\beta} \,,
\label{cpol}
\ee
where $n_{j}(r_{V})$ is the size distribution of
non-spherical dust grains of the type $j$, $r_{V}$ radius of equivolume sphere
(for infinite circular cylinders the
particle radius $r_{\rm cyl}$ is used).
The superscripts TM and TE denote two cases of
orientation of the electric vector of the incident radiation relative to the
particle axis \cite{bh83}.
The average polarization cross sections $\overline{C}_{{\rm pol},\,j}$
depend on the alignment function $f(\xi, \beta)$ with the alignment
parameter $\xi$. Here, ${\beta}$ is the precession-cone angle for
the angular momentum $\vec{J}$ which spins around
the direction of the magnetic field $\vec{B}$,
${\varphi}$ the spin angle,
${\omega}$  the precession angle (see Fig.~\ref{f-sky}).
In general, the particles are assumed to be partially aligned:
the major axis of the particle rotates in the spinning plane which is
perpendicular to the angular
momentum which spins (processes) around the direction of the
magnetic field. The angle between the line of sight and the magnetic
field is $\Omega$ ($0^\circ \leq \Omega \leq 90^\circ$).

The cross sections
$C^{\rm TM, \, TE}_{\rm ext} = \pi r_{V}^2 \, Q^{\rm TM, \, TE}_{\rm ext}$
in Eq.~(\ref{cpol}) can be calculated by using the light scattering theory
for non-spherical particles.
To facilitate the calculations, particles of simple shapes
are usually considered (see Table~\ref{p-mod}).
\setlongtables
\begin{longtable}{lll}
\caption{\small Models used for interpretation of \is polarization.}\label{p-mod} \\
\hline\noalign{\smallskip}
Author(s) (year) reference & Dust grains & Alignment \\
                           &             & (angle) \\
\noalign{\smallskip}\hline\noalign{\smallskip}
\endfirsthead
\noalign{\smallskip}\hline
\endfoot
\caption{(Continued.)} \\
\hline\noalign{\smallskip}
Author(s) (year) reference & Dust grains & Alignment \\
                           &             & (angle) \\
\noalign{\smallskip}\hline\noalign{\smallskip}
\endhead
Wilson (1960) \cite{w60}          & infinite cylinders$^*$ & PDG \\
                                  &                        & ($\Omega=30\degr,\,90\degr$)\\
Greenberg et al. (1963-8)  & infinite cylinders, & PF, PDG \\
\cite{gre68,glwl63,gs66}                                  & homogeneous spheroids$^{**}$ &  \\
Rogers and Martin  (1979)   & homogeneous spheroids & PF \\
\cite{rm79}                                   &                       & ($\alpha=90\degr$)\\
Hong and Greenberg (1980)    & silicate core--ice mantle & IDG, PDG \\
\cite{hg80}                                  & infinite cylinders &  \\
Mathis (1979, 1986)  &  infinite cylinders$^{***}$ & PDG$^{+}$ \\
\cite{m79,m86}                                  & (spheroids)                & ($\Omega=90\degr$)\\
Onaka (1980) \cite{ona80}                & core--mantle spheroids & PF \\
                                  &                        & ($\alpha=90\degr$)\\
Vaidya et al. (1984) \cite{va84}         & homogeneous  & PF \\
                                  & spheroids$^{****}$     &  ($\alpha=45\degr,\,90\degr$)\\
Voshchinnikov  et al.  & silicate core--ice mantle & IDG, PDG  \\
(1986, 1989) \cite{vii86,v89}                                  & infinite cylinders &     \\
Mishchenko  (1991)  \cite{mishch91}      & homogeneous spheroids & IDG, PDG \\
Wolff et al. (1993) \cite{wcm93}         & homogeneous,  coated & PDG$^{+}$  \\
                                  & and composite infinite   & ($\Omega=90\degr$) \\
                                  & cylinders, homogeneous  &  \\
                                  & spheroids  &  \\
Kim and Martin (1994) \cite{km94}        & infinite cylinders  & PDG  \\
                                  &                     &  ($\Omega=90\degr$) \\
Kim and Martin (1995) \cite{km95}        & homogeneous spheroids & PF, PDG  \\
                                  &                       & ($\alpha,\,\Omega=90\degr$) \\
Matsumura and Seki (1996)     & homogeneous spheroids  & PF  \\
\cite{ms96}                                  & and ellipsoids & ($\alpha=90\degr$)\\
Li and Greenberg  (1997)      & silicate core--organic & PDG \\
\cite{lig97}                                  & mantle finite cylinders & ($\Omega=90\degr$)\\
Vaidya  et al. (2007) \cite{vgs07}       & silicate spheroids with & PF ($\alpha=45\degr,$ \\
                                  & graphite inclusions  & $60\degr,\,90\degr$)\\
Wurm and Schnaiter (2002)   & dust aggregates   & PF \\
\cite{ws02}                                   & consisting of 4--64 & ($\alpha=90\degr$)\\
                                  & monomers & \\
Voshchinnikov et al. (1990-            & homogeneous spheroids & IDG, PDG \\
2010)  \cite{vd08,v90,vf93,dvi10}       &  &                               \\
Draine and Fraisse (2009)     & oblate spheroids & PDG$^{++}$ \\
\cite{df09}                                  &                  & ($\Omega=90\degr$)
\end{longtable}
\protect\vspace*{-0.4cm}
{\small
\noindent PF --- picket fence orientation \\
PDG --- perfect Davis--Greenstein (2D) orientation\\
IDG --- imperfect Davis--Greenstein orientation \\
$\alpha,\, (\Omega)$ --- angle between the line of sight and the direction of
grain alignment in the case of PF (PDG) orientation \\
$^*$ in Rayleigh-Gans approximation;  \\
$^{**}$ in Rayleigh approximation;  \\
$^{***}$ efficiency factors tabulated by Wickramasinghe \cite{wicr73} were taken;
polarization for prolate spheroids was computed as for cylinders;   \\
$^{****}$ the figures of Asano \cite{as79} were used;  \\
$^{+}$ large silicate grains are  assumed to be perfectly aligned
(see Eq.~(\ref{mathis})); \\
$^{++}$ only grains with sizes $r>r_{\rm cut}$ are assumed to be perfectly
aligned (see Eq.~(\ref{drfr})). \\
}
\vspace*{5pt}

\noindent Early models dealt with homogeneous infinitely long circular cylinders
\cite{glwl63,gs66,m79,m86}\footnote{Another simplification was
the use of the wavelength-independent refractive index.}.
This is the simplest model of non-spherical
particles. Solution to the light scattering problem for infinite cylinders
was obtained by the separation of variables method 
in the cylindrical coordinate system \cite{lindg66}.
Later, more advanced models with silicate core-ice
mantle cylindrical particles based on the solution from
\cite{shah70}
were developed \cite{hg80,vii86,v89}.
The progress in the light scattering theory allowed one to apply
the model of homogeneous prolate and oblate spheroids of different size
and shape for calculations of the polarizing efficiency, visual and
UV polarization (see, e.g., \cite{rm79,km95,dvi10,df09}).
Spheroidal particles are characterized
by the  aspect ratio $a/b$ where $a$ and $b$ are the major and
minor semiaxes.
The optical properties of spheroids can be found
by using different technique. The most popular methods
widely applied in astronomical modelling are
the separation of variables method \cite{vf93,ay75},
the T-matrix method \cite{mishch91} and the discrete dipole
approximation \cite{d00}.
Comparison of the methods and benchmark results are given
in \cite{hovetal96,vihf00}\footnote{see also
Database of Optical Properties (DOP): {\tt http://www.astro.spbu.ru/DOP}}.
More complicated non-spherical particles
(coated spheroids, ellipsoidal particles, composite spheroids)
were considered so far for illustrative calculations
\cite{ona80,ms96,vgs07}.


The polarization cross sections must be averaged
over rotations taking into account an alignment mechanism.
According to standard concepts \cite{w03,kru03},
the alignment of interstellar grains may be magnetic or radiative.
A very popular alignment mechanism is the magnetic alignment
(Davis--Greenstein (DG) type orientation \cite{dg51})
based on the paramagnetic relaxation of grain material containing about
one percent of iron impurities.
For the imperfect Davis--Greenstein (IDG) orientation,  the alignment function
${f}(\xi, \beta)$ can be written as \cite{hg80,v89}
\be
{f}(\xi, \beta) = \frac{\xi \sin \beta}{(\xi^2 \cos^2 \beta  + \sin^2 \beta)^{3/2}}.
\label{idg} \ee
The parameter $\xi$ depends on the particle size $r_V$, the imaginary part of
the grain magnetic susceptibility
$\chi''$ ($=\varkappa \omega_{\rm d} /T_{\rm d}$, where $\omega_{\rm d}$ is
the angular velocity
of grain), gas  density $n_{\rm g}$, the strength of magnetic field $B$
and dust ($T_{\rm d}$) and gas ($T_{\rm g}$) temperatures
\be
\xi^2  = \frac{r_V +\delta_0 (T_{\rm d}/T_{\rm g})}{r_V +\delta_0},
\,\,\,\,\,\,
{\rm where}
\,\,\,\,\,\,
\delta_0^{\rm IDG} = 8.23\,10^{23} \frac{\varkappa B^2}{n_{\rm g} T_{\rm g}^{1/2} T_{\rm d}}\,\mkm.
\label{xi} \ee
If the grains are not aligned $\xi=1$ and ${f}(\xi, \beta)=\sin \beta$;
in the case of the perfect rotational orientation  $\xi=0$.
Unfortunately,
only a limited number of models
including the combined particle size/shape/orientation  analysis
have been developed (see Table~\ref{p-mod}).

For simplicity, many investigators assumed that the direction of the magnetic
field (direction of grain alignment) was perpendicular to the line of sight, i.e.,
$\alpha,\,\Omega=90\degr$ (e.g., \cite{m86,km95,df09}).
Frequently, non-rotating particles of the same orientation
are  considered (see Table~\ref{p-mod}). In this case, called
the ``picket fence'' (PF) orientation, there are no integrals over the angles
$\varphi, \, \omega$ and  $\beta$ in Eq.~(\ref{cpol}).
The polarization degree is proportional to the polarization cross-section
$P  \propto C_{\rm pol} =
1/2 [C^{\rm TM}_{\rm ext}(\Omega) -C^{\rm TE}_{\rm ext}(\Omega)]$, where
$\Omega = \alpha$ and $f(\xi, \beta)=\delta(\alpha)$\footnote{$\delta(z)$
is the Dirac delta function.}.
The dichroic polarization efficiency is defined
by the ratio of the polarization  cross-section (factor) to the
extinction  one
\be
\left( \frac{P}{\tau} \right)_{\rm PF} =
 \frac{C_{\rm pol}}{C_{\rm ext}} =
 \frac{C^{\rm TM}_{\rm ext}-C^{\rm TE}_{\rm ext}}
{C^{\rm TM}_{\rm ext}+C^{\rm TE}_{\rm ext}} \cdot 100\% =
 \frac{Q^{\rm TM}_{\rm ext}-Q^{\rm TE}_{\rm ext}}
{Q^{\rm TM}_{\rm ext}+Q^{\rm TE}_{\rm ext}} \cdot 100\%.
\label{ppf}
\ee
%
A more complicated case is the perfect rotational (2D)  orientation
(or perfect Davis--Greenstein orientation, PDG)
when the major axis of a non-spherical particle always lies in the same plane.
For the 2D orientation,  integration is performed over the spin
angle $\varphi$ only and $f(\xi, \beta)=\delta(\beta)$.

\begin{figure}[htb]
\centerline{
\resizebox{7cm}{!}{\includegraphics{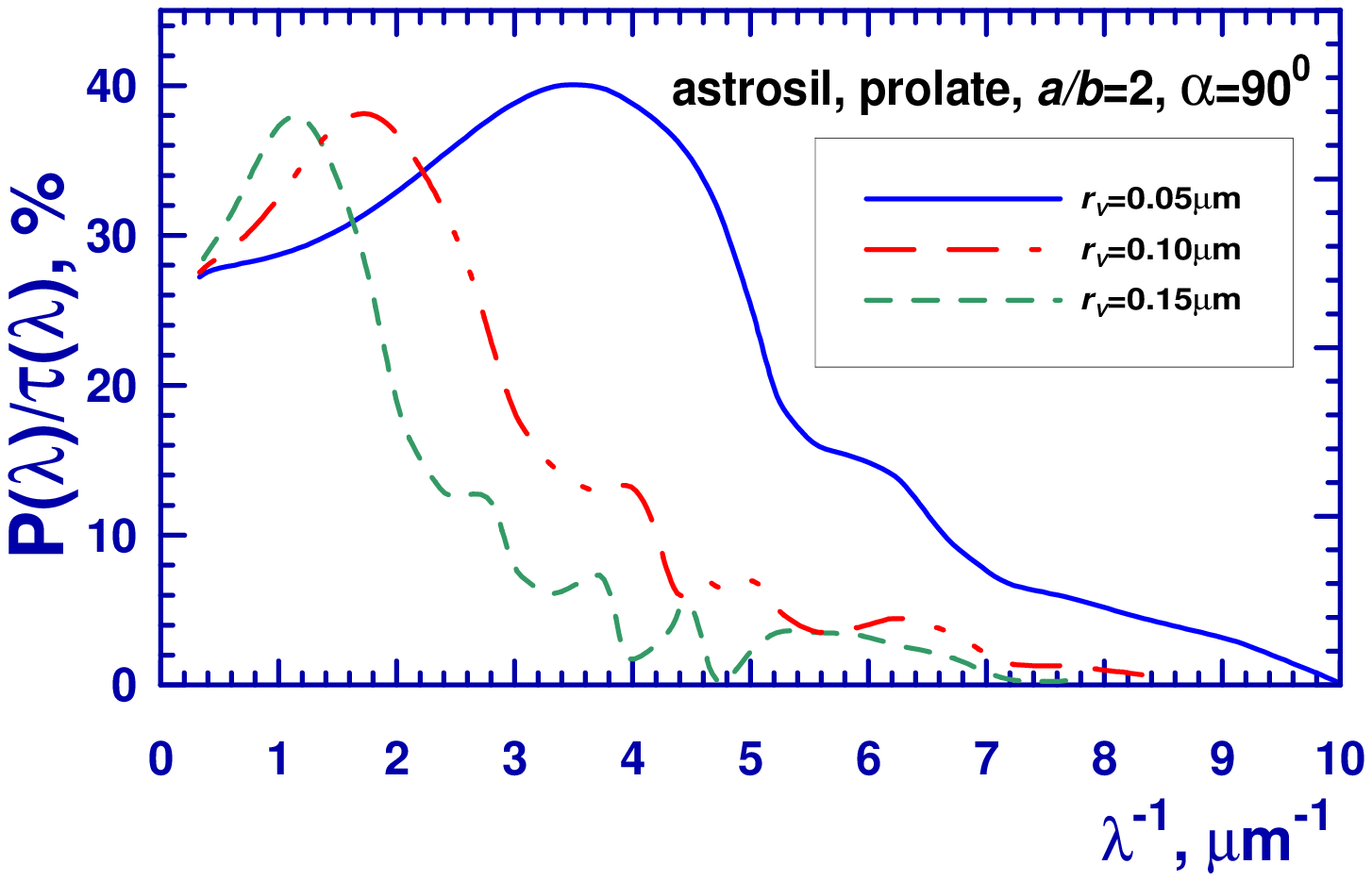}}
\resizebox{7cm}{!}{\includegraphics{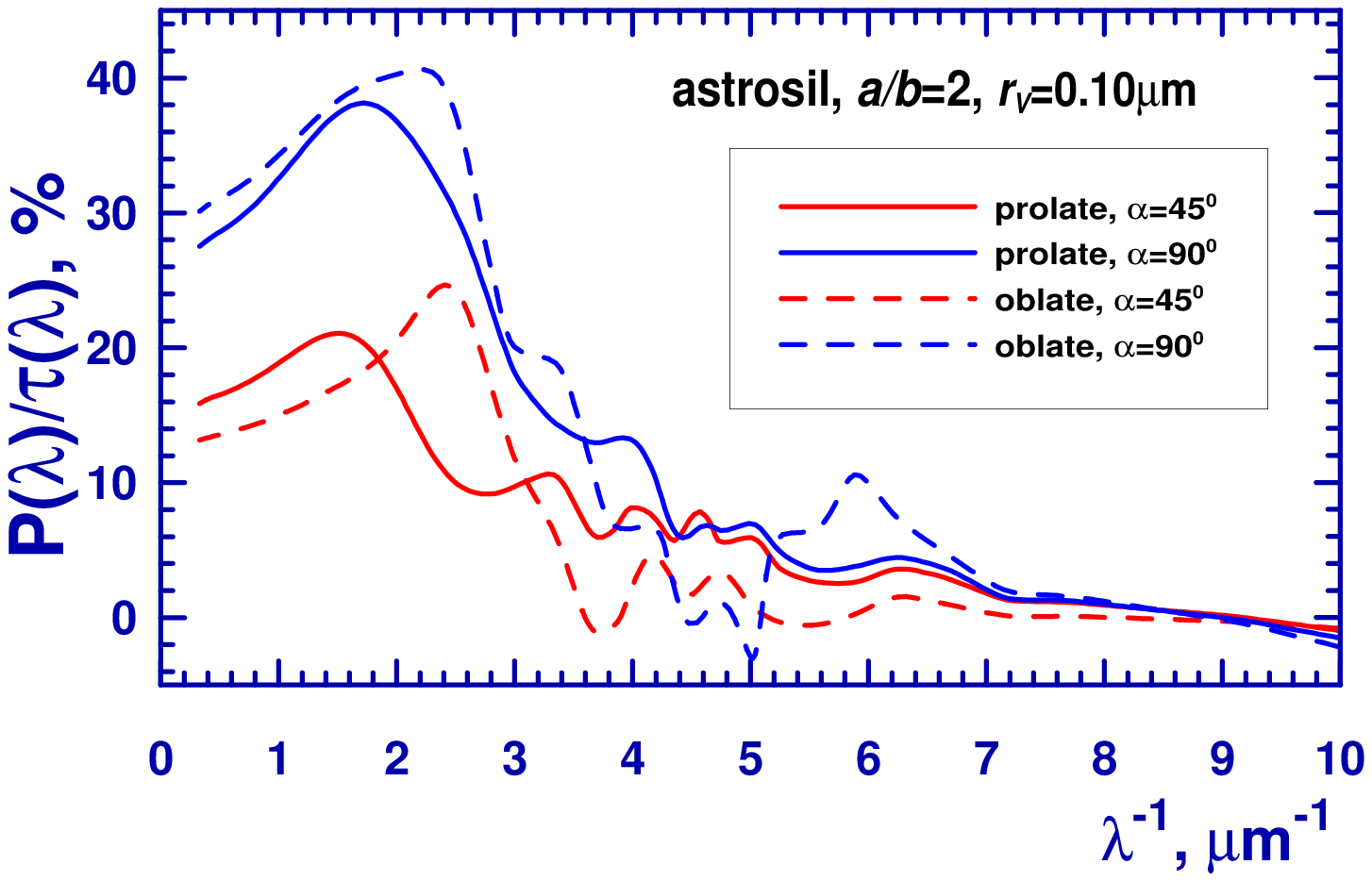}}
}
\caption{Wavelength dependence of the polarization efficiency
for homogeneous spheroids consisting of astrosil for the PF orientation
(see Eq.~(\ref{ppf})).
The effect of variations of the particle size (left panel),
type and orientation (right panel) is
illustrated. Adapted from \cite{v04}.}
\label{f-ptaupf}
\end{figure}

\begin{figure}[htb]
\centerline{
\resizebox{7cm}{!}{\includegraphics{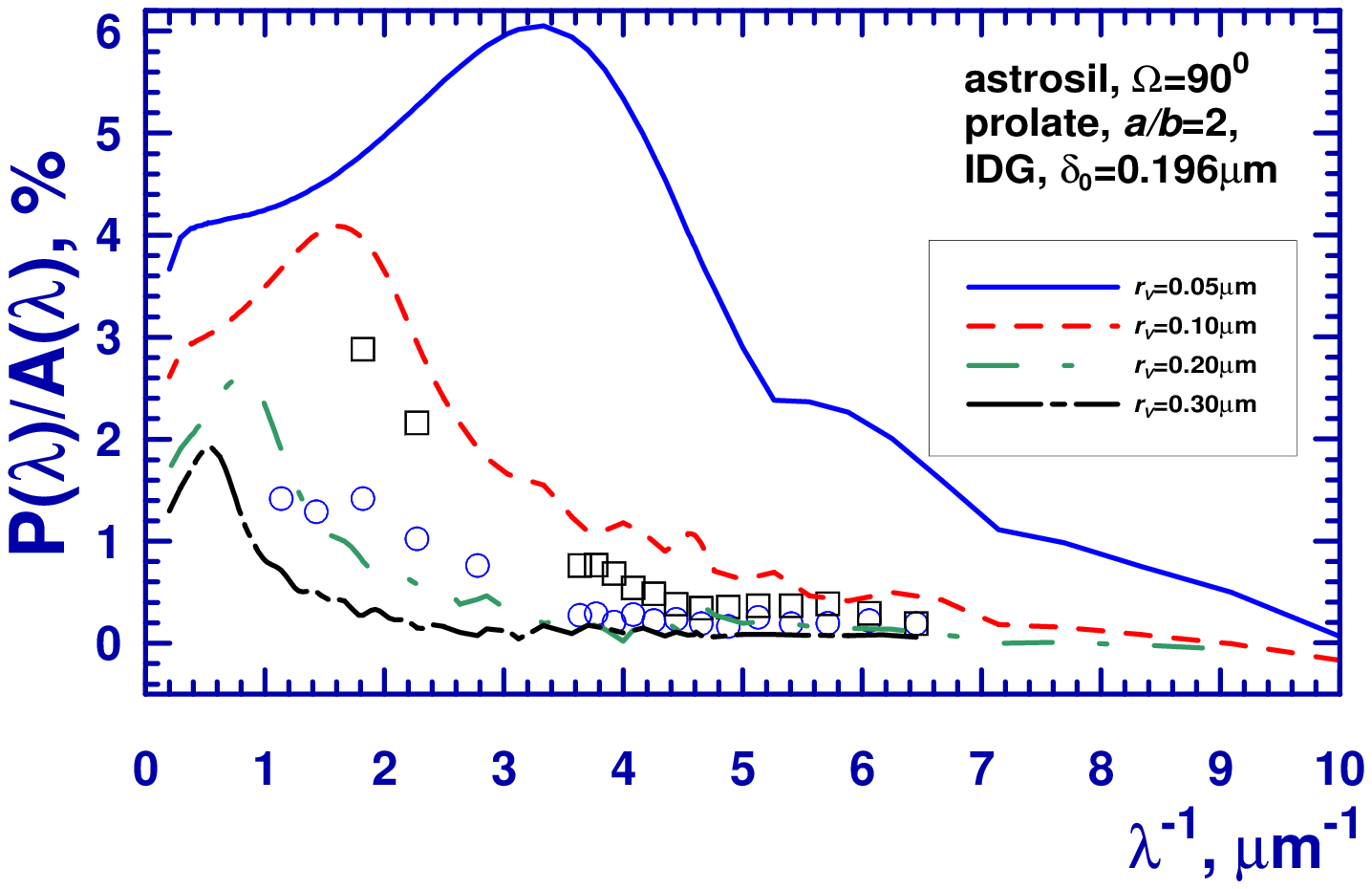}}
\resizebox{7cm}{!}{\includegraphics{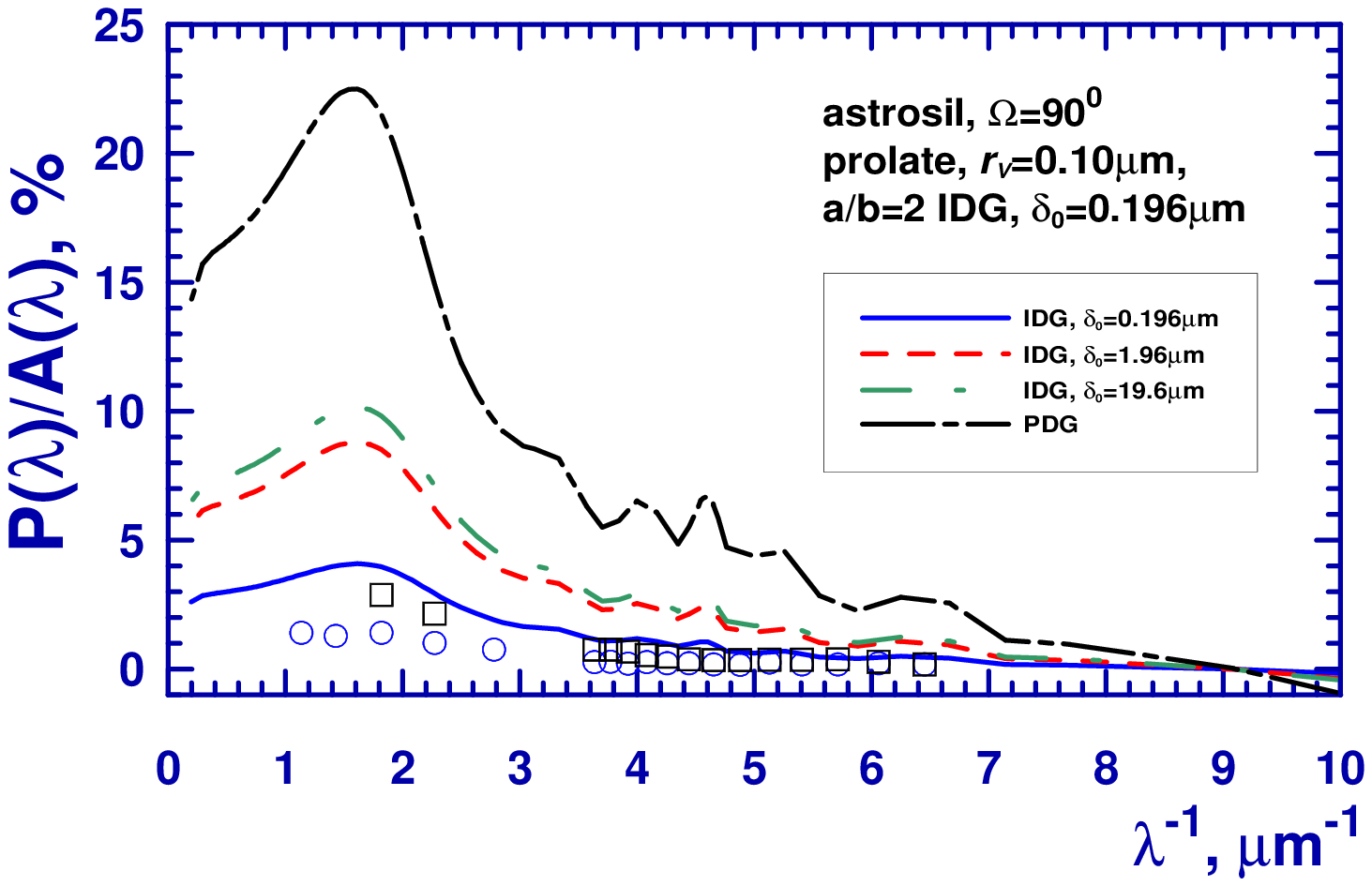}}
}
\caption
{Wavelength dependence of the polarization   efficiency  for
homogeneous rotating spheroidal particles of the astronomical silicate.
The effect of variations of particle size (left panel), and degree
of alignment  (right panel) is illustrated.
The open circles and squares show the observational data
for stars  HD~24263 and HD~99264, respectively.
Adapted from \cite{vd08}.}
\label{f-ptidg}
\end{figure}

Polarization produced by perfectly aligned particles
is much larger than that observed (cf. Figs.~\ref{f-ptaupf}
and \ref{f-ptidg} with Eq.~(\ref{pebv})).
Nevertheless, the models with the PF or PDG orientation
are useful for investigations of the
normalized polarization (the Serkowski curves) as the wavelength
dependence of polarization is only slightly  influenced by
the particle refractive index, size or shape (cf. left panels in
Figs.~\ref{f-ptaupf} and \ref{f-ptidg}).
As a crude approximation of the polarization efficiency for particles
with the IDG orientation, the following relation is used (see, e.g., \cite{rm79})
$$
\left( \frac{P}{\tau} \right)_{\rm IDG} = {\cal R}\, \sin^2 \Omega\,
\left( \frac{P}{\tau} \right)_{\rm PF},
$$
where ${\cal R}=1/2(3\langle \cos^2 \beta \rangle -1)$
is the Rayleigh reduction factor \cite{gre68,rl99} and $\langle \rangle$
denotes the ensemble average. It should be emphasized that application of
this approximation as well as the Rayleigh reduction factor can lead
to misinterpretation of observational curves $P(\lambda)$.

In the case of the IDG mechanism, smaller grains are
aligned better than  larger grains
(see Eq.~(\ref{xi})). However, the models
with an opposite type of orientation of small and large particles have been
also suggested.
Mathis \cite{m86} assumed that rotating  silicate grains were
perfectly aligned if they contain at least one super-paramagnetic inclusion.
Carbonaceous grains and silicate grains without inclusions are randomly oriented
in space (3D orientation). The probability of perfect alignment is
\be
{f}(r_V,r_V')= 1- \exp (-r_V/r_V')^3.
\label{mathis}\ee
Draine and Fraisse \cite{df09} considered the model of
silicate and amorphous carbon spheroids with
randomly oriented small particles and perfectly aligned large particles.
In this case the alignment function is size dependent
\begin{equation}
{f}(\beta,r_V) = \left\{
\begin{array}{ll}
 \sin \beta & {\rm for}\ \ r_V \le r_{V,\,\rm cut}, \\
 \delta(\beta) & {\rm for}\ \ r_V > r_{V,\,\rm cut},
\end{array}  \right.
\label{drfr}
\end{equation}
where $r_{V,\rm cut}$ is a cut-off parameter.

Computations made by Das et al. \cite{dvi10} (see their Fig.~2) demonstrate
that the observational data can be fitted by using the models with
different alignment functions
(e.g., given by Eqs.~(\ref{idg}), (\ref{mathis}) or (\ref{drfr})), especially if
a more complex size distribution function as discussed in Sect.~\ref{i-hom}
is  chosen. However, this complicates the model.
To avoid the models with many parameters
new ideas about the nature of polarizing grains and physics of grain alignment
should be included.  

The DG mechanism of the paramagnetic relaxation requires a stronger
magnetic field than average Galactic magnetic field ($\sim$3 -- 5\,$\mu$G;
\cite{hc05}).
Because of this problem, it has been suggested that the polarizing
grains contain small clusters of iron, iron sulfides, or iron oxides
with super-paramagnetic or ferromagnetic properties
\cite{js67}. This leads to an enhancement of the
imaginary part of the
magnetic susceptibility of grain material $\chi''$ {\rm by a factor 10 -- 100}
and alignment can occur through the DG mechanism.
This scenario is supported by laboratory experiments
 \cite{dj07,bel09}.
A significant enhancement of $\chi''$ is also possible in mixed
MgO/FeO/SiO grains \cite{dul78} or in H$_2$O ice mantle grains
containing magnetite (Fe$_3$O$_4$) precipitates \cite{sor94,sor95}.

Another possibility to align interstellar grains is the radiative torque
alignment (RAT alignment). It arises from an azimuthal asymmetry of
the light scattering by non-spherical particles.
Magnetic inclusions can enhance RAT alignment \cite{lh08}.
The theory of RAT alignment is well developed  \cite{lh07}.
Recent observations of interstellar polarization in the vicinity of
luminous stars
\cite{ap10,mat11,and11} have been used for confirmation of the RAT
alignment mechanism.
However, the discussed models are phenomenological, they are not based on
correct light scattering calculations of interstellar polarization.
One of the reasons is that the alignment function for the RAT mechanism is unknown.
 Another reason is a requirement of advanced light scattering methods
because fast rotation can only occur for grains of very specific (helical)
shape \cite{lh07,dgs79}.
This is highly improbable from the point of view of grain growth
in the interstellar medium.

Since both magnetic alignment and radiative alignment depend on iron
inclusions, we can expect that polarization and/or polarization efficiency
should increase with the growth of iron fraction in dust grains.
This idea was investigated by Voshchinnikov et al. \cite{vhpd12} by using
 available data on interstellar polarization and element
abundances  previously compiled in \cite{vh10}.
It was suggested that the interstellar polarization was probably related to
the amount of iron in dust grains.
Assuming that all silicon and all magnesium
are embedded into amorphous silicates of olivine composition
(Mg$_{2x}$Fe$_{2-2x}$SiO$_4$,
where $x = [{\rm {Mg}/{H}}]_{\rm d}/(2 [{\rm {Si}/{H}}]_{\rm d})$
as is a part of iron.
The remaining part of Fe can be found as
\be
\left [{\rm {Fe(rest)}/{H}} \right ]_{\rm d}  =
\left [{\rm {Fe}/{H}} \right ]_{\rm d}  -
(2\,\left [{\rm {Si}/{H}} \right ]_{\rm d} -
\left [{\rm {Mg}/{H}} \right ]_{\rm d}).   \label{fer}
\ee
\begin{figure}[htb]
\centerline{
\resizebox{7cm}{!}{\includegraphics{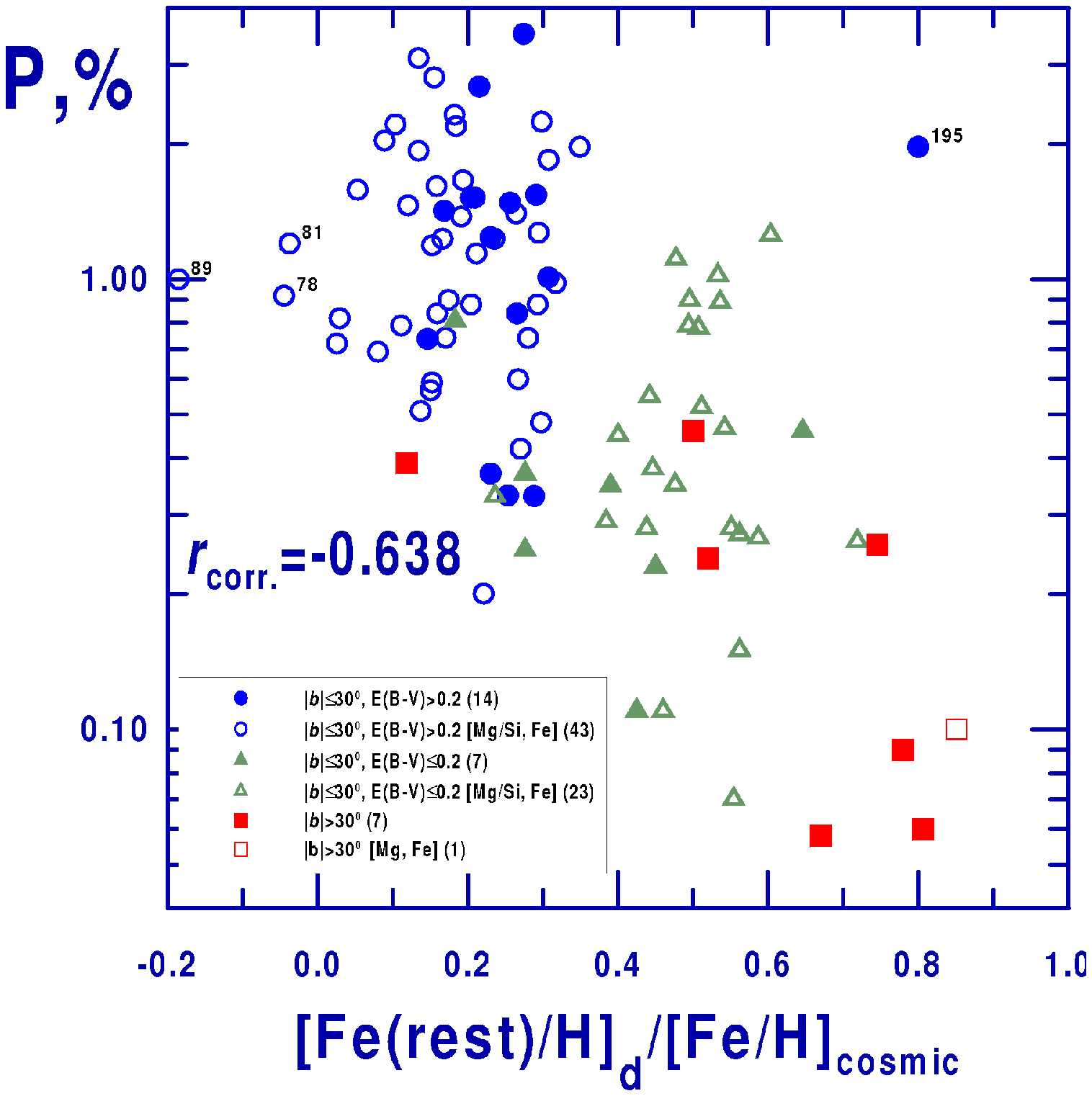}}
\resizebox{7cm}{!}{\includegraphics{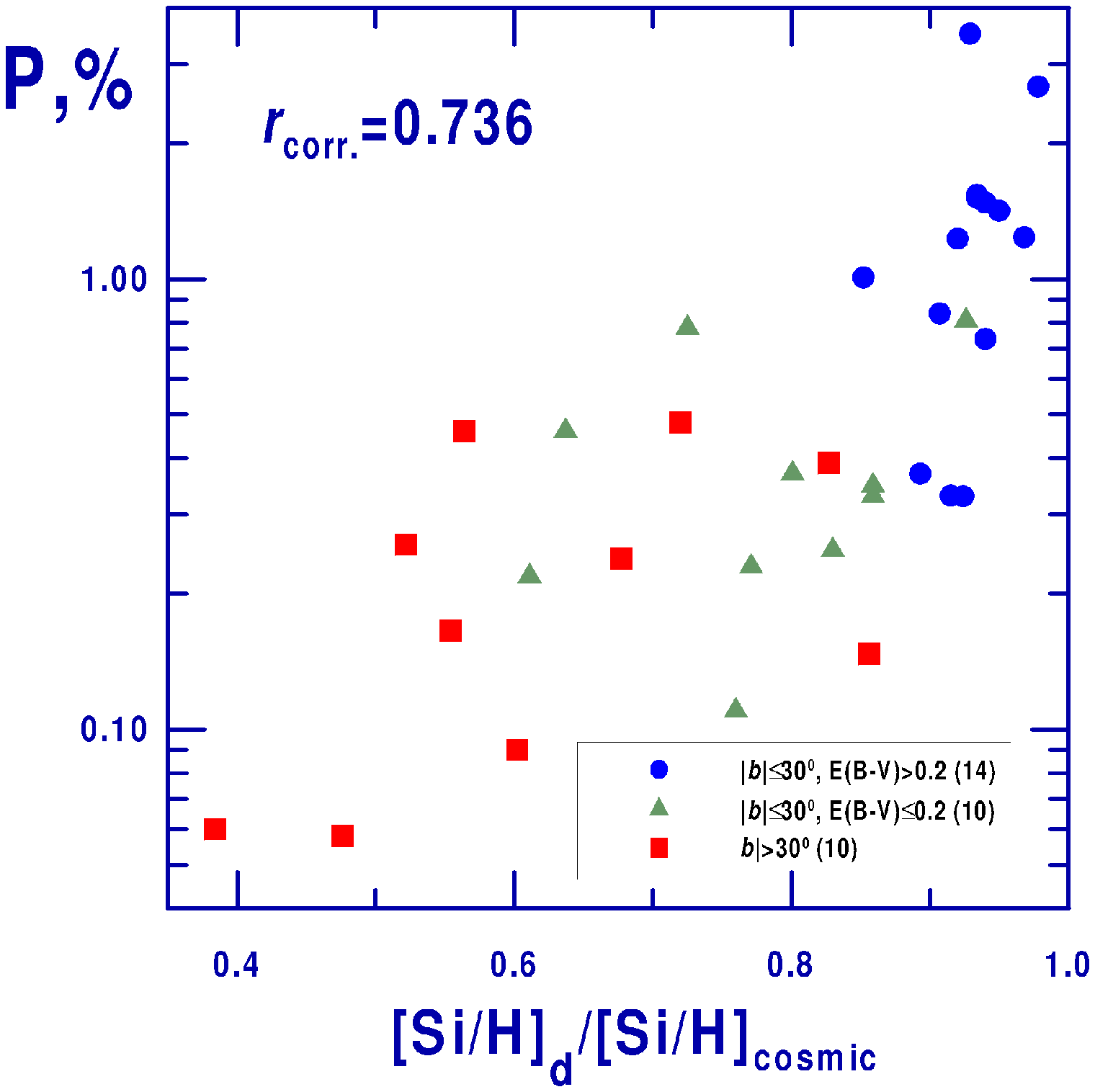}}
}
\caption{Interstellar polarization in dependence on remaining dust phase abundance
of Fe as given by Eq.~(\ref{fer}) 
(left panel) and dust phase abundance of silicon (right panel).
Halo stars with $|b| > 30\degr$ and disk stars with $|b|\leq 30\degr$
and low ($E(B-V) \leq 0.2$) and high ($E(B-V) > 0.2$) reddening
are shown with different symbols.
Number of stars is indicated in parentheses in the legend.
Adapted from \cite{vhpd12}.}
\label{f-pfe}
\end{figure}
As indicated in Fig.~\ref{f-pfe} (left panel), there is a negative correlation
between the polarization degree $P$ and the amount of remaining iron.
This is inconsistent with the common suggestion about
the great role of iron-rich grains in the production of polarization.
Since $P$ is proportional to the column density of polarizing grains,
we can conclude that the increase of the iron content in non-silicate grains
does not enhance polarization.

Because in calculating $\left [{\rm {Fe(rest)}/{H}} \right ]_{\rm d}$
we removed all Si and Mg and a part of Fe from the dust phase,
we expect a positive correlation between the  polarization and the
abundances of the eliminated elements.
There is only a weak correlation between
$P$ and  $\left [{\rm {Fe}/{H}} \right ]_{\rm d}$ or
$\left [{\rm {Mg}/{H}} \right ]_{\rm d}$ (see \cite{vhpd12} for more
discussion) and a strong correlation between
$P$ and  $\left [{\rm {Si}/{H}} \right ]_{\rm d}$ (Fig.~\ref{f-pfe},
right panel).
Therefore, it can be established that polarization is more likely produced
by silicates.
These findings are evidence in favour of
the assumption of Mathis \cite{m86} that only the silicate grains are
aligned and contribute to the observed polarization, while
the carbonaceous grains are either spherical or randomly aligned.
Another verification of this suggestion is the absence of
any correlation between the polarization efficiency $P/E(B-V)$ or $P/A_V$
and dust phase abundances of elements
(see \cite{vhpd12}).
This is because dust grains of all types (silicate,
carbonaceous, iron-rich, etc.) contribute to the observed extinction, 
while only the silicates
seems to be responsible for the observed polarization.
Thus, the  absence of correlation between
$R_V$ and  
$\lambda_{\max}$ (see discussion in Sect.~\ref{p-obs})
can be easily understood. These discoveries can be
explained if the silicate grains aligned by the radiative mechanism
are mainly responsible for the observed interstellar linear
polarization\footnote{Polarization in  IR features also
supports the idea of separate populations of polarizing
(silicate) and non-polarizing (carbonaceous) grains. This follows
from the observed polarization of silicate features at
10\,$\,\mkm$ and 18\,$\,\mkm$ and the lack of polarization
in the 3.4\,$\,\mkm$ hydrocarbon feature (see \cite{ha05,chair06}
and references therein).}.


Analysing models presented in Table~\ref{p-mod}, it is possible to say that
the major part of models includes perfect grain alignment and one angle of
alignment and merely a few of them with the IDG orientation can be used for
interpretation of observations of individual stars.  Indeed, the
previous modelling of \is polarization was mainly focused on the explanation
of the average wavelength dependence (Serkowski curve) \cite{m79,m86,km94,km95}.
Only Li and Greenberg~\cite{lg98} applied their
model of coated cylinders to explain the normalized polarization curve
in the direction of  HD~210121 and Das et al. \cite{dvi10} interpreted
\is extinction and polarization observations of seven stars using
a mixture of carbonaceous and silicate spheroids.
This fact causes  deep dissatisfaction because a great
amount of observations of interstellar polarization in different areas
exists and continuously grows.

\subsection{Interpretation: dust grains and magnetic field in
the Taurus dark cloud}\label{tdc}

In this section we present the
quantitative interpretation of observations of \is polarization
for a group of stars\footnote{More detailed discussion can be found
in \cite{vy12}.}.
Our model of spheroidal grains with imperfect alignment \cite{vd08,dvi10}
is supplemented by the subroutine
calculating three parameters of the  Serkowski curve.
We also assume that polarization is  mainly produced by silicate grains
and the degree of alignment of carbonaceous grains is small
(see  discussion in previous section).

As an example we refer to the Taurus dark cloud (TDC) --- the complex of \is
clouds where active star formation is in progress. This complex lies
sufficiently far from the Galactic plane ($b \approx - 15$\degr) at a
distance of $\sim$~140~pc and comprise several dozens dark nebulae, clouds and
clumps \cite{kgw08,dob05}. It suffers negligible foreground and background
extinction    \cite{str80,pcl02} and is the subject of numerous
investigations of molecular gas \cite{schs84,guw84,crut85,heyer87,gold08}.

The polarimetric observations of several hundreds stars in the region around
the Taurus dark cloud complex have been performed at  visual and red
wavelengths \cite{heyer87,mon84,good90}
and in the  H and K bands \cite{gwl95,tam87,good92,chap11}.
The obtained polarization maps give an insight into the
structure of the plane-of-the-sky component of \is
magnetic field  but are of little use for studies of
grain properties and alignment. To find these latter,
the wavelength dependence of polarization must be involved.
Whittet et al. \cite{wetal01} presented the polarimetric
and photometric observations of 27 stars in a wide  spectral range
in the TDC ($l \approx$ 170\degr $\div$ 176\degr,
$b \approx$ --10\degr $\div$   --17\degr)
and calculated the fit parameters of the Serkowski curve $P_{\max}$, $\lambda_{\max}$ and
$K$ as well as  the values of $R_V$.
Our initial analysis of these data is based on the two-component model of
Messenger et al. \cite{mess97} and Whittet et al. \cite{wetal04}
(see also discussion of Table~\ref{p-mess} in Sect.~\ref{p-obs}).
It made possible to form two groups of stars with relatively
uniform distribution of positional angles of polarization:
cloud 1 (14 stars, $\theta_{\rm gal.}=145\degr - 175\degr$) and
cloud 2 (13 stars, $\theta_{\rm gal.}=2\degr - 40\degr$).

Firstly, we should concentrate on stars in cloud 1 located
inside or behind the ``diffuse--screen'' component of the TDC.
These stars are distributed around Heiles Cloud 2
($l \approx$ 174\fdegr4,  $b \approx$ --13\fdegr4, area 15.8~pc$^2$ \cite{gold08})
--- a dense condensation containing TMC-1 (Taurus molecular cloud).
\begin{figure}[htb]
\centerline{
\resizebox{7cm}{!}{\includegraphics{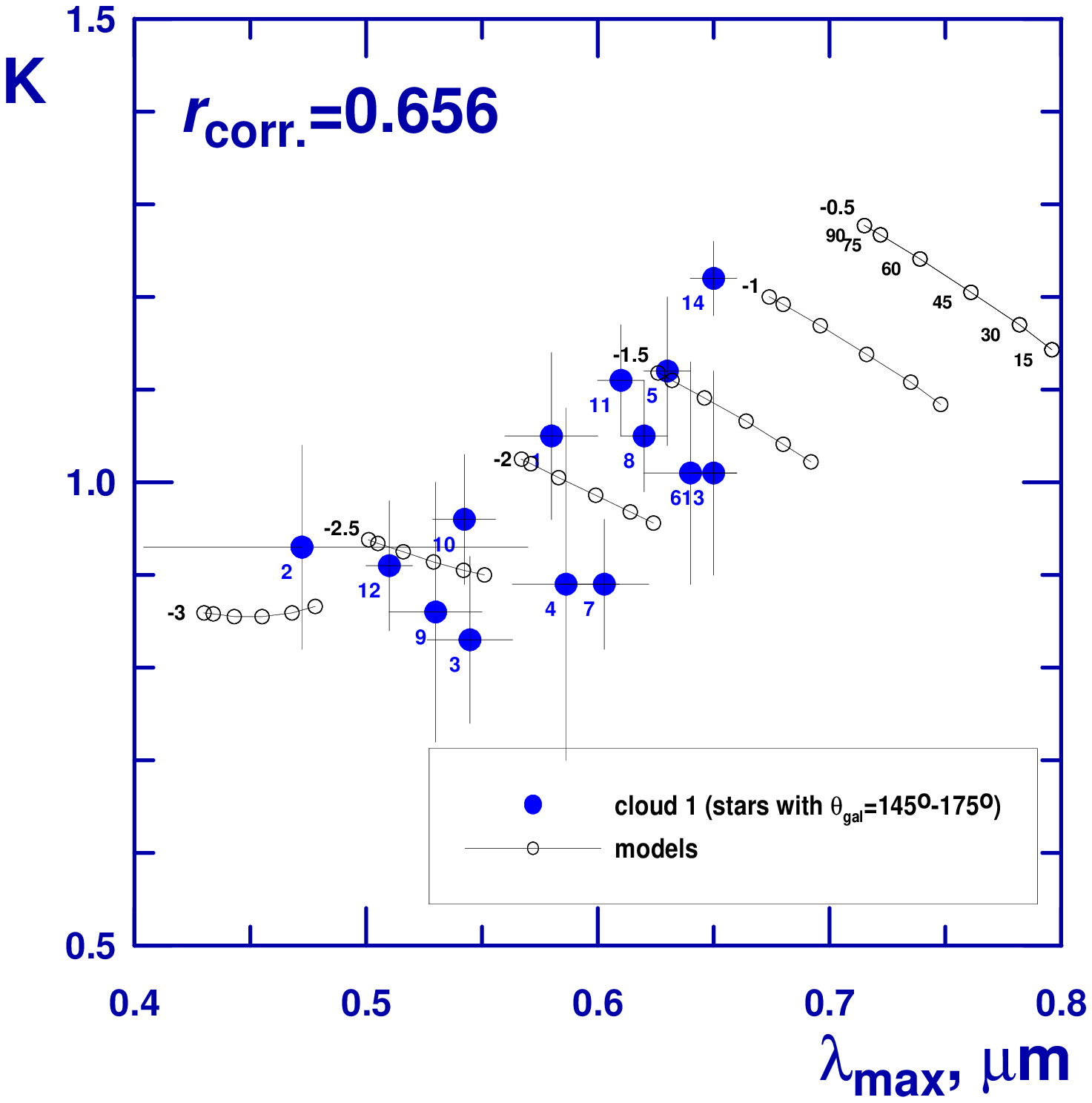}}
\resizebox{7cm}{!}{\includegraphics{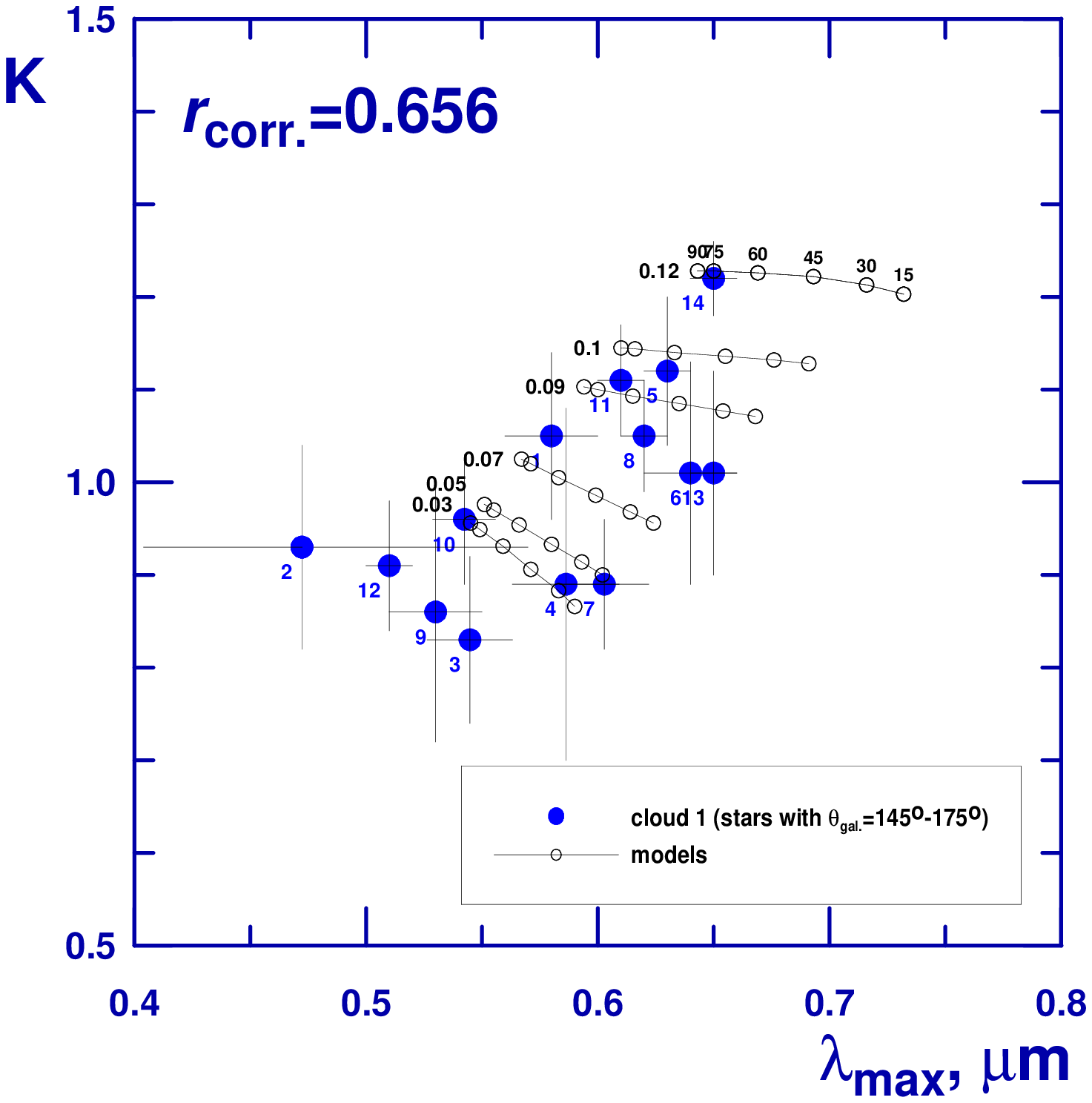}}
}
\caption{The coefficients $K$ of the Serkowski curve (\ref{serkk})
in dependence on the wavelength of maximum polarization $\lambda_{\max}$
for 14 stars in cloud 1 in Taurus with the similar positional angles
of polarization. The numbers of stars correspond to the increasing HD numbers
of stars in Fig.~\ref{klm}, i.e. 1 = HD~28225, \dots, 14 = HD~283879.
The Pearson correlation coefficient between
$K$ and  $\lambda_{\max}$ is given. Open circles with line show
model calculations with different alignment angles
$\Omega=15\degr(15\degr)90\degr$. The values of $\Omega$ increase from right
to left as marked for the top model. Left panel illustrates the effect
of variations of the power index $q$ in the power-law size distribution
($q$ varies from --0.5 to --3). Right panel illustrates the effect
of variations of the lower cut-off $r_{V, \min}$  in the power-law size
distribution ($r_{V, \min}$ varies from $0.03\,\mkm$ to $0.12\,\mkm$).
Other model parameters are:
prolate spheroids, $a/b = 3$, $r_{V, \max}=0.35\,\mkm$, $r_{V, \min}=0.07\,\mkm$
(left panel), $q=-2$ (right panel).
}
\label{f-t1}
\end{figure}
\begin{figure}[htb]
\centerline{
\resizebox{7cm}{!}{\includegraphics{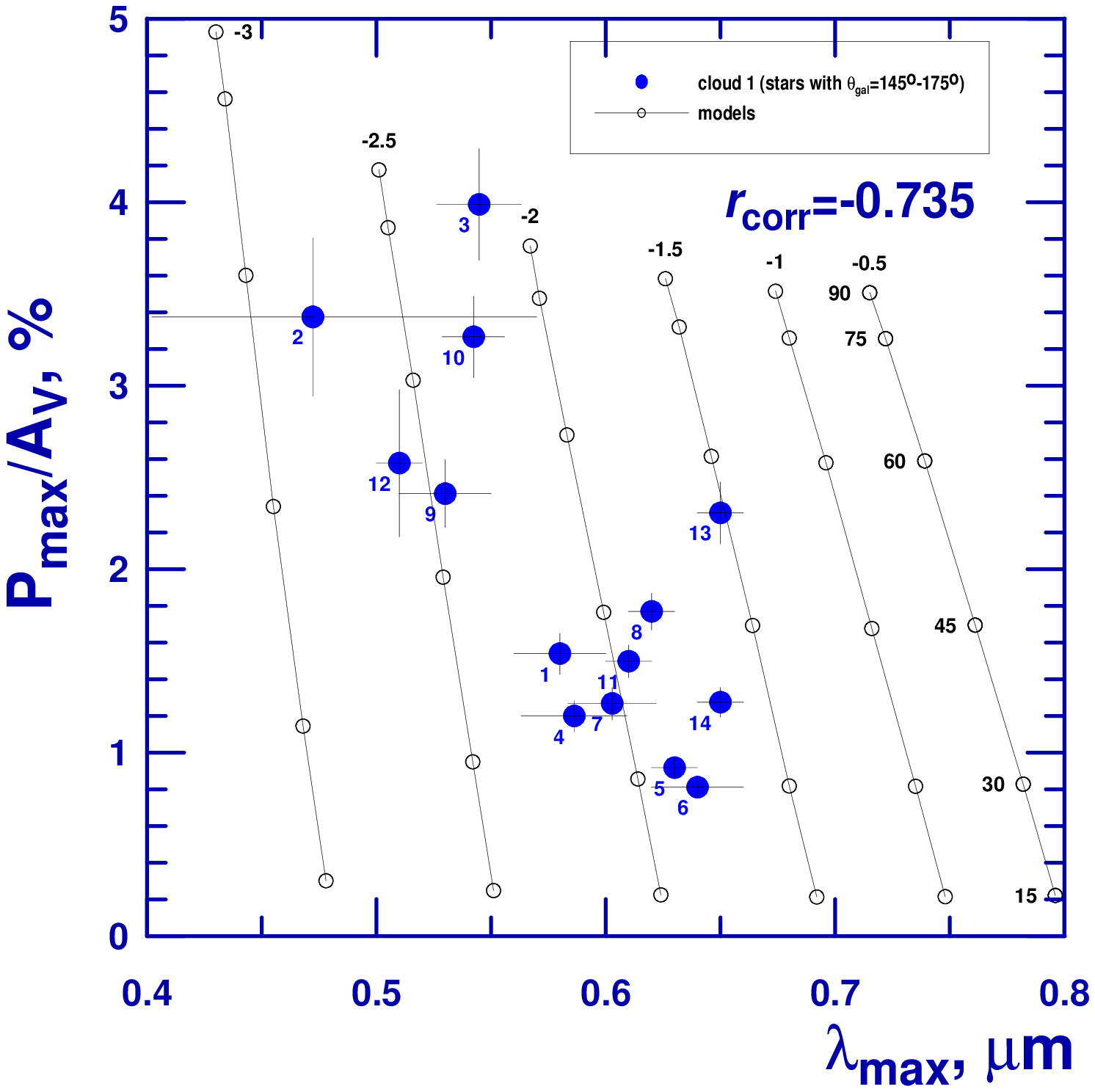}}
\resizebox{7cm}{!}{\includegraphics{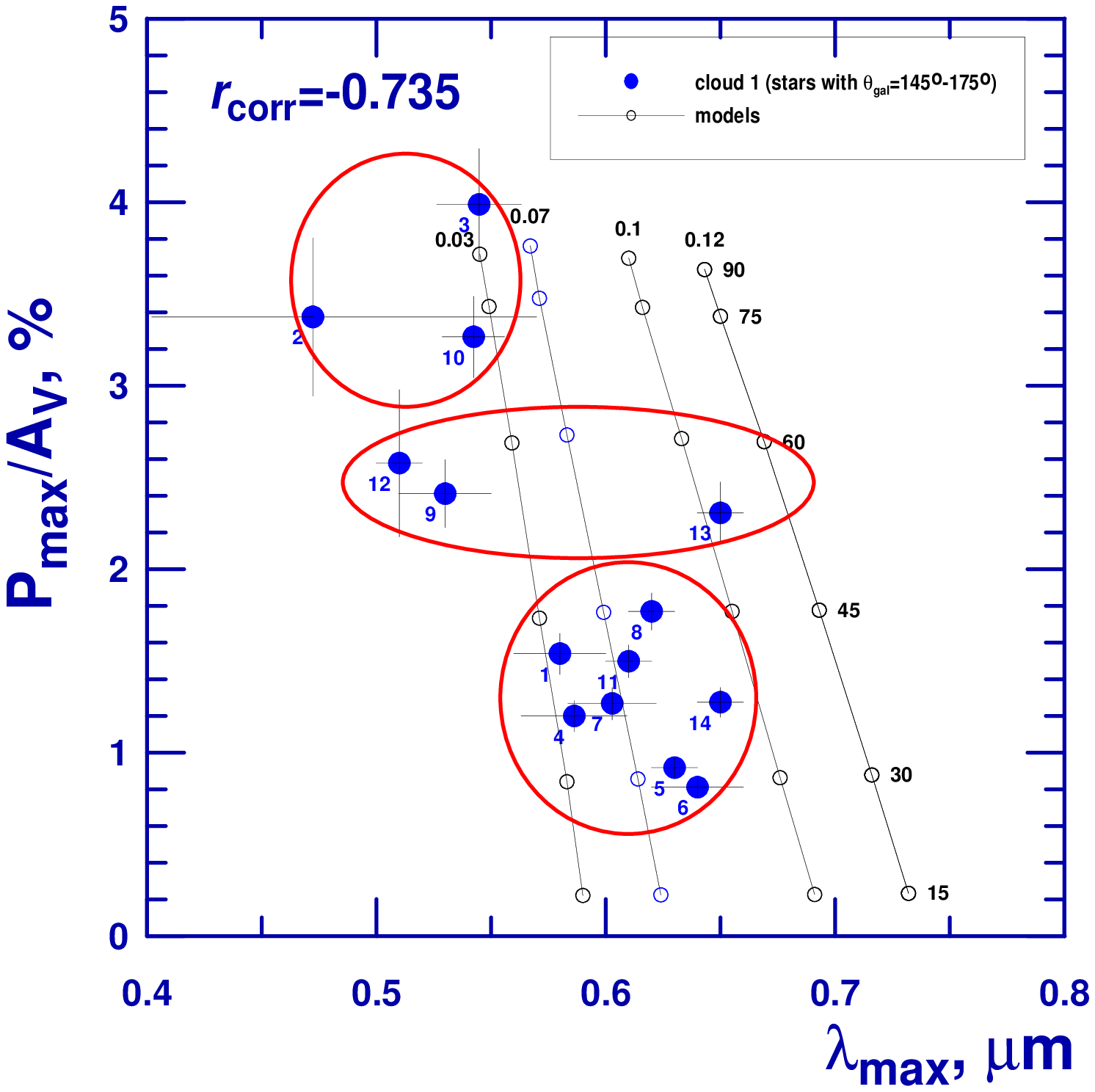}}
}
\caption{The polarizing efficiency $P_{\max}/A_V$
in dependence on the wavelength of maximum polarization $\lambda_{\max}$
for 14 stars in cloud 1 in Taurus.
The Pearson correlation coefficient between
$P_{\max}/A_V$  and  $\lambda_{\max}$ is given. Open circles with
line show  model calculations with different alignment angles
$\Omega=15\degr(15\degr)90\degr$. The values of $\Omega$ increase from bottom
to top as marked for the right model. Left panel illustrates the effect
of variations of the power index $q$ in the power-law size distribution
($q$ varies from --0.5 to --3). Right panel illustrates the effect
of variations of the lower cut-off $r_{V, \min}$  in the power-law size
distribution ($r_{V, \min}$ varies from $0.03\,\mkm$ to $0.12\,\mkm$).
Other model parameters are:
prolate spheroids, $a/b = 3$, $r_{V, \max}=0.35\,\mkm$, $r_{V, \min}=0.07\,\mkm$
(left panel), $q=-2$ (right panel).
}
\label{f-t2}
\end{figure}
Observational data for stars in the cloud 1 plotted in Figs.~\ref{f-t1}, \ref{f-t2}
indicate  positive correlation between $K$ and $\lambda_{\max}$
(the width of the polarization curve $W$ decreases with increasing
$\lambda_{\max}$) and negative correlation between the polarization efficiency
$P_{\max}/A_V$ and $\lambda_{\max}$.
The first dependence is well known (see Fig.~\ref{klm} and discussion in
\cite{wetal92}). Its qualitative explanation is a systematic reduction in the relative
number of small, aligned grains in regions of high $\lambda_{\max}$
\cite{wetal01,mcw99}. The sole
quantitative modelling 
was
attempted by Aannestad and Greenberg
\cite{ag83}. However, to reduce computational
efforts they ruled out the integration over the angle $\omega$
in Eqs.~(\ref{cpol}) which led to wrong results for $W$ \cite{v89}.
A systematic trend toward smaller polarizing efficiency
for larger $\lambda_{\max}$ was detected for nearby stars closely located
on the sky \cite{wetal01,wetal94,vy10}.
It has been interpreted as a result of the decrease of the angle $\Omega$
between the direction of the magnetic field and the line of sight
\cite{v89}.

\begin{figure}[htb]
\centerline{\resizebox{9.3cm}{!}{\includegraphics{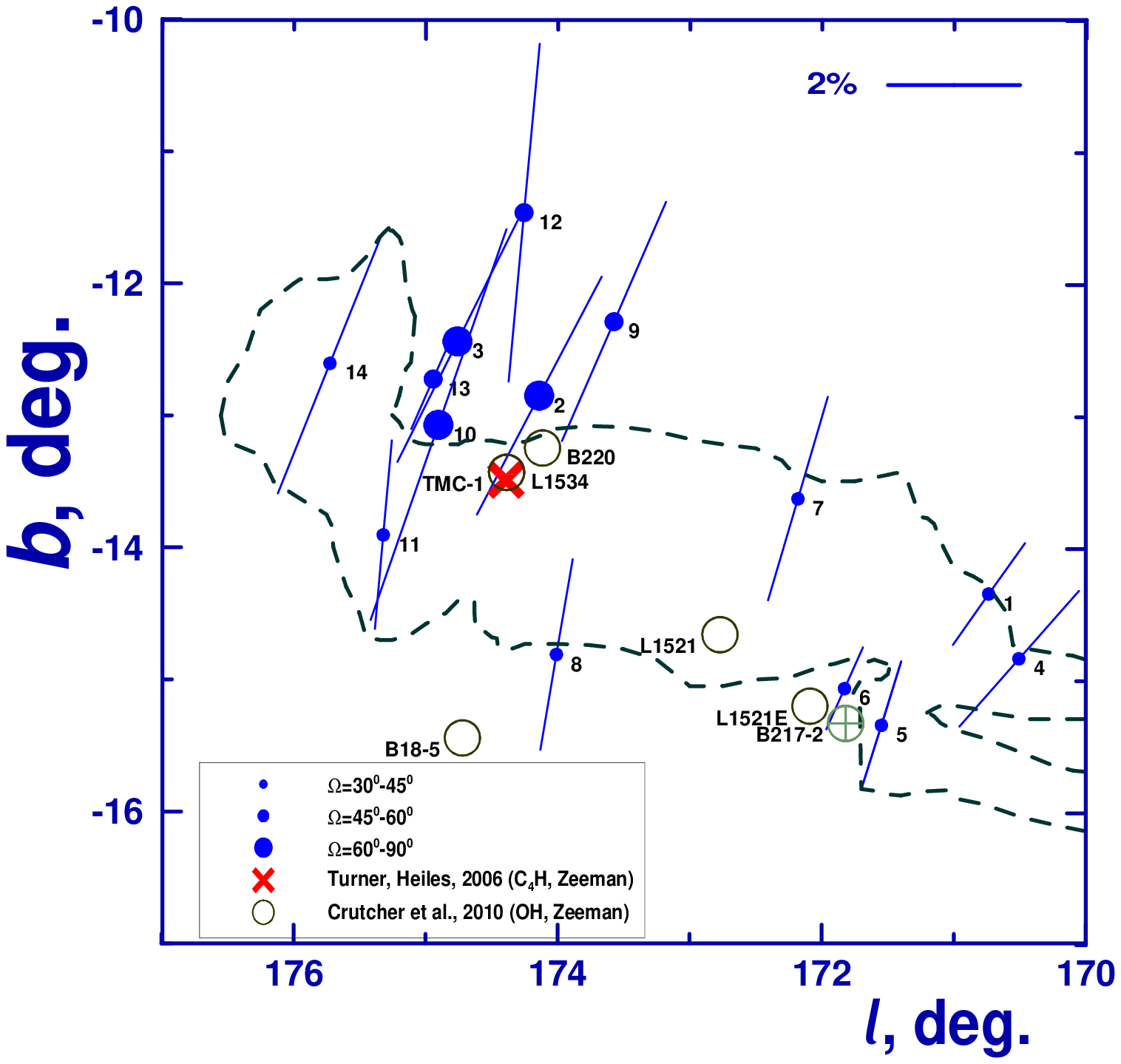}}}
\caption
{Linear polarization map of the cloud 1 in Taurus
containing 14 stars with  similar positional angles.
The lengths of the lines are proportional to the percent polarization.
The sizes of the circles are proportional to the  alignment angle
$\Omega$. The dashed contour represents the regions with different
visual extinction: $1\fm6 < A_V < 19\fm6$ inside contour and
 $0\fm4 < A_V < 1\fm6$ outside contour \cite{pcl02,lee06}.
The cross corresponds to the clump in TMC-1 where Turner and
 Heiles \cite{th06} have searched the C$_4$H Zeeman effect.
The open circles show the positions of dark clouds where
the OH Zeeman effect has been observed  \cite{crut10}.
}
\label{f-t3}
\end{figure}

The results of theoretical modelling are shown in Figs.~\ref{f-t1}, \ref{f-t2}
by open circles connected with solid line. The particle shape was fixed: prolate
spheroids with the aspect ratio $a/b = 3$.
Under assumptions made
the main parameters of our model influencing the polarization are:
the lower and upper cut-offs $r_{V, \min}$
and $r_{V, \max}$ and the power index $q$ in the power-law size distribution
for silicate particles  and the degree (parameter $\delta_0^{\rm IDG}$,
see Eq.~(\ref{xi})) and direction (angle $\Omega$) of alignment.
Left and right panels in Fig.~\ref{f-t1} illustrate the variations
of the index $q$ and lower cut-off $r_{V, \min}$, respectively.
The rise of these parameters  may be associated with the growth of
dust grains by coagulation ($q$) or accretion ($r_{V, \min}$).
In both cases the mean size of grains is bigger at the right upper corner
of Fig.~\ref{f-t1} in comparison with its left bottom corner.
It is interesting
that the stars NN 1, 5, 6, 8, 11, 14 located  at the right upper corner
apparently are embedded in Heiles Cloud 2 or situated at its boundary
(see Fig.~\ref{f-t3}). Note also that variations of alignment
parameters $\delta_0^{\rm IDG}$ and $\Omega$ only
do not allow explaining
the observed correlation between $K$ and $\lambda_{\max}$.

The opposite situation occurs with
the correlation between the polarization efficiency $P_{\max}/A_V$ and
$\lambda_{\max}$ (Fig.~\ref{f-t2}). In this case
to explain both high and low values of $P_{\max}/A_V$  it is not
sufficient to change the grain size only, variations of
parameters $\delta_0^{\rm IDG}$ and $\Omega$ must be taken into account.
The theoretical points plotted in Fig.~\ref{f-t2} were obtained
for $\delta_0^{\rm IDG} = 3\,\mu$m. Smaller values of $\delta_0^{\rm IDG}$
do not reproduce the data for directions with high polarization efficiency
(stars NN 2, 3, 10).
It is evident that the obtained
grain parameters are model-dependent (e.g., it is possible to use particles
of another shape). However,
with our model of \is dust the trends
in variations of the grain size, shape, alignment can be determined.

We attribute the variations of $P_{\max}/A_V$
in Fig.~\ref{f-t2} to the changes in the direction
of magnetic field (direction of grain alignment).
Stars can be divided into three groups with close values of $\Omega$
(see right panel in Fig.~\ref{f-t2}). These groups of stars are shown
in Fig.~\ref{f-t3} by  filled circles of different sizes.
This Figure gives the positions and polarization of stars in cloud 1 and
approximate contours of Heiles Cloud 2.
It is intriguing that all six stars with the larger values of $\Omega$
(the magnetic field is mainly perpendicular to the line of sight)
are closely located on the sky outside of Heiles Cloud 2.
Other eight stars where the magnetic field is significantly tilted to
the line of sight are situated near the boundary of the cloud.
An indirect support to our interpretation is  provided by Zeeman observations.
Turner and Heiles \cite{th06} obtained an upper limit
for magnetic field toward the cold dense TMC-1 cyanopolyyne peak core
$B_{\|}=B\,\cos \Omega = 14.5 \pm 14 \,\mu$G.
A possible reason for this result is that the magnetic field is directed
close to the plane of the sky.
Crutcher et al. \cite{crut10} compiled the Zeeman data for many dark
clouds. Several of them  located near or inside Heiles Cloud 2
are marked as large open circles in Fig.~\ref{f-t3}.
In almost all cases the data are within 1$\sigma$ error.
Only for B217-2 (close to the positions of stars NN 5, 6)
the component of the magnetic field parallel to the
line of sight has been detected ($B_{\|}= 13.5 \pm 3.7\,\mu$G).

The structure and evolution of magnetic field in Heiles Cloud 2
have been discussed by Heyer et al. \cite{heyer87} and
Tamura et al. \cite{tam87}. It was suggested that the contraction
and formation of the cloud
from the placental Taurus dark cloud with homogeneous magnetic structure
occurred along the magnetic field.
The gas motions in Heiles Cloud 2 can be described in terms of
a rotating ring with the rotation axis coinciding with
the magnetic field \cite{schs84}. Our findings infer that the
magnetic field is tilted with respect to the line of sight
at the boundaries of Heiles Cloud 2, i.e.,
magnetic field has a spindel-like
structure. Apparently, the area disturbed by contraction of Heiles Cloud 2
terminates at the place where
the magnetic field is almost perpendicular to the line of sight
(stars NN 2, 3, 9, 10, 12, 13).

\section{Conclusions}

All modern investigations of cosmic dust in
protoplanetary disks, dense clouds, distant galaxies are
based on  the modelling of ``classical'' observations of the
\is extinction and polarization. Our examination shows that the
interpretation of these basic observations still remains incomplete.
New generation models
should include a consideration of
interstellar abundances in given directions
and accurate treatment of
light scattering by physically feasible particles
with realistic alignment.

\section*{Acknowledgments}
I am grateful to organizers F. Borghese, M.A. Iat\`\i\, and R. Saija
for inviting me to present this review.
I am thankful to the referees for useful comments,
L. Cambr\'esy and P. Padoan for sending the extinction
map in numerical form, I.S. Yakovlev and H.K. Das for assistance in
calculations and to V.B. Il'in for production Serkowski fitting subroutine
and careful reading of manuscript.
The work was partly supported by the grant RFBR 10-02-00593a
of the Russian Federation.


\end{document}